\documentclass[review]{elsarticle}
\usepackage{color}
\usepackage{epstopdf}
\usepackage{amsmath}
\usepackage{lineno,hyperref}
\usepackage{color}
\usepackage[english]{babel}
\usepackage{latexsym}
\usepackage{graphicx}
\usepackage{babel}
\usepackage{amsmath}
\usepackage{amssymb}
\usepackage{subfloat}
\usepackage{subfigure}
\usepackage{morefloats}
\usepackage{varioref}
\biboptions{numbers,sort&compress}

\labelformat{equation}{Eq.~(#1)}
\labelformat{figure}{Figure~#1}
\labelformat{table}{Table~#1}
\linenumbers
\makeindex

\journal{Chemical Engineering Research \& Design}

\begin{document}
\nolinenumbers

\begin{frontmatter}

\title{Flow characterisation and power consumption in an inline high shear rotor-stator  mixer using CFD}
\author[man]{Vipin Michael \corref{correspondingauthor}}
\cortext[correspondingauthor]{Corresponding author}
\ead{vipin.michael@manchester.ac.uk}
\author[ncl]{Umair Ahmed\corref{null0}}
\author[naj]{Mahmoud Assad\corref{null0}}
\author[man]{Robert Prosser\corref{null0}}
\author[ul]{Adam Kowalski\corref{null0}}

\address[man]{School of Engineering, University of Manchester, Manchester, M13 9PL, UK}
\address[ncl]{School of Engineering, Newcastle University, Newcastle upon Tyne, NE17RU}
\address[naj]{Mechanical Engineering Department, An-Najah National University, Nablus, Palestine}
\address[ul]{Unilever R\&D, Port Sunlight Laboratory, Quarry Road East, Bebington, \\ Wirral CH63 3JW, UK}

\begin{abstract}
The aim of this paper is two-fold: (1) to provide a detailed investigation of the turbulent flow in an inline high-shear rotor stator mixer; (2) to provide a comparison of two different classes of turbulence models and solution methods currently available. The widely used multiple reference frame (MRF) method is contrasted against a more recently developed sliding mesh method. The sliding mesh algorithm accounts for rotation of the blades and is able to capture the transient effects arising from the rotor-stator interaction. The choice of turbulence model is shown to have a significant impact, with second moment closures able to capture best the hydrodynamics. With an appropriate choice of turbulence model and solution algorithm, we thus demonstrate the capacity of CFD to provide accurate and computationally cost effective characteristic power curve predictions.
\end{abstract}

\end{frontmatter}

\section{Introduction}
The mixing process plays a significant role in improving the homogeneity and quality of a wide range of products in the fast moving consumer goods industries (i.e. pharmaceutical, biomedical, agricultural, cosmetic, health care and food processing). Inline rotor-stator mixers are widely used in processing due to their high efficiency and their capacity to accelerate the mixing process by providing a focussed delivery of energy \cite{Jasinska2015}. However, the high energy dissipation rates and short residence times within the mixer limits current understanding of the fluid dynamics within these devices and consequently their relationship to overall mixer performance \cite{Obeng2004}. 

Rotor-stator mixers consist of high speed rotors surrounded by close fitting stator screens. The typical tip speeds during operation range from $10-50\textrm{m/s}$, and the gaps between the rotor and stator range between $100-3000\mu\textrm{m}$ \cite{Utomo2009},  generating high shear rates in the rotor-stator gap ranging from $20,000\textrm{s}^{-1}-100,000 \textrm{s}^{-1}$ \cite{Obeng2004}. The high kinetic energy imparted to the fluid by the rotating blades is mainly dissipated local to the stator screen;  the high rate of energy dissipation makes such devices advantageous for physical processes such as mixing, dispersion, dissolution, emulsification and de-agglomeration \cite{Kowalski2011}.

The power curve is one of the main tools used to characterise high shear mixers, since power consumption is intimately linked to the overall energy dissipation and thus provides a comparative basis for the mixer performance. The power curve is also useful for scale up calculations \cite{Ozcan2011}. Recently, efforts based on experimental methods have been made to characterise and predict the power consumption of inline Silverson mixers \cite{Cooke2008,Cooke2012,Ozcan2011,Kowalski2011,Hall2011}. However, investigations of the detailed flow structures and mixing within these devices are still limited. Baldyga et al. \cite{Baldyga2007, Baldyga2008} and Jasi\'nska et al. \cite{Jasinska2014, Jasinska2015} have carried out CFD simulations of an inline Silverson 150/250 MS in-line mixer focussing on estimating the product yield during chemical reaction, distribution of particle aggregates and droplet size distributions. The details of the fluid dynamics within the mixer were limited;  simulations were carried out using the standard $k-\epsilon$ turbulence closure via a multiple reference frame (MRF) model. Qualitative agreement was found between the experimental and simulation results although details of the transient flow (due to periodic passing of the blades in the front of the stator cavities) were lost due to the inability of MRF to simulate the rotor rotation. In addition, standard eddy viscosity closures are not sensitive to fluid rotation and streamline curvature, and hence their use limits the predictive capability of CFD simulations in these mixers \cite{22}. In Michael et al \cite{michael2017cfd} Unsteady Reynolds Averaged Navier-Stokes (URANS) simulations on a sliding mesh were performed for the fluid dynamics, linking the $k-\omega$ SST turbulence model to the population balance equations. Drop dispersion and non-Newtonian rheology of dense emulsions in the mixer was investigated using a combined CFD-PBM approach. 

This paper builds upon these earlier CFD investigations by presenting a detailed investigation into the turbulent flow dynamics arising in the inline Silverson 150/250MS mixer. A sliding mesh algorithm is used to capture the interaction between the rotating and stationary volumes within the mixer. Turbulence is modelled using both rotation-curvature compensated eddy viscosity models (EVMs), and second moment closures (Reynolds stress models or RSMs). The latter class of models are able to account for rotation and curvature effects in a systematic manner, due to the presence of exact production terms containing mean flow gradients and system rotation, but they come at a higher computational cost. The ability to predict power consumption, strongly swirling turbulent flow, and mixing, using both EVMs and RSM  closures forms the major output of this work. 

The paper is organised as follows: in the next section we briefly describe the test configuration investigated. Section \ref{sec:nmerical_code_sec} outlines the numerical procedure and in section \ref{sec:Turb_models} the different turbulence models are described. Results are presented and discussed in section \ref{sec:results}, with the conclusions summarised in the last section.

\section{Test configuration}
\begin{figure}[htbp]
\centering
\includegraphics[scale=0.55]{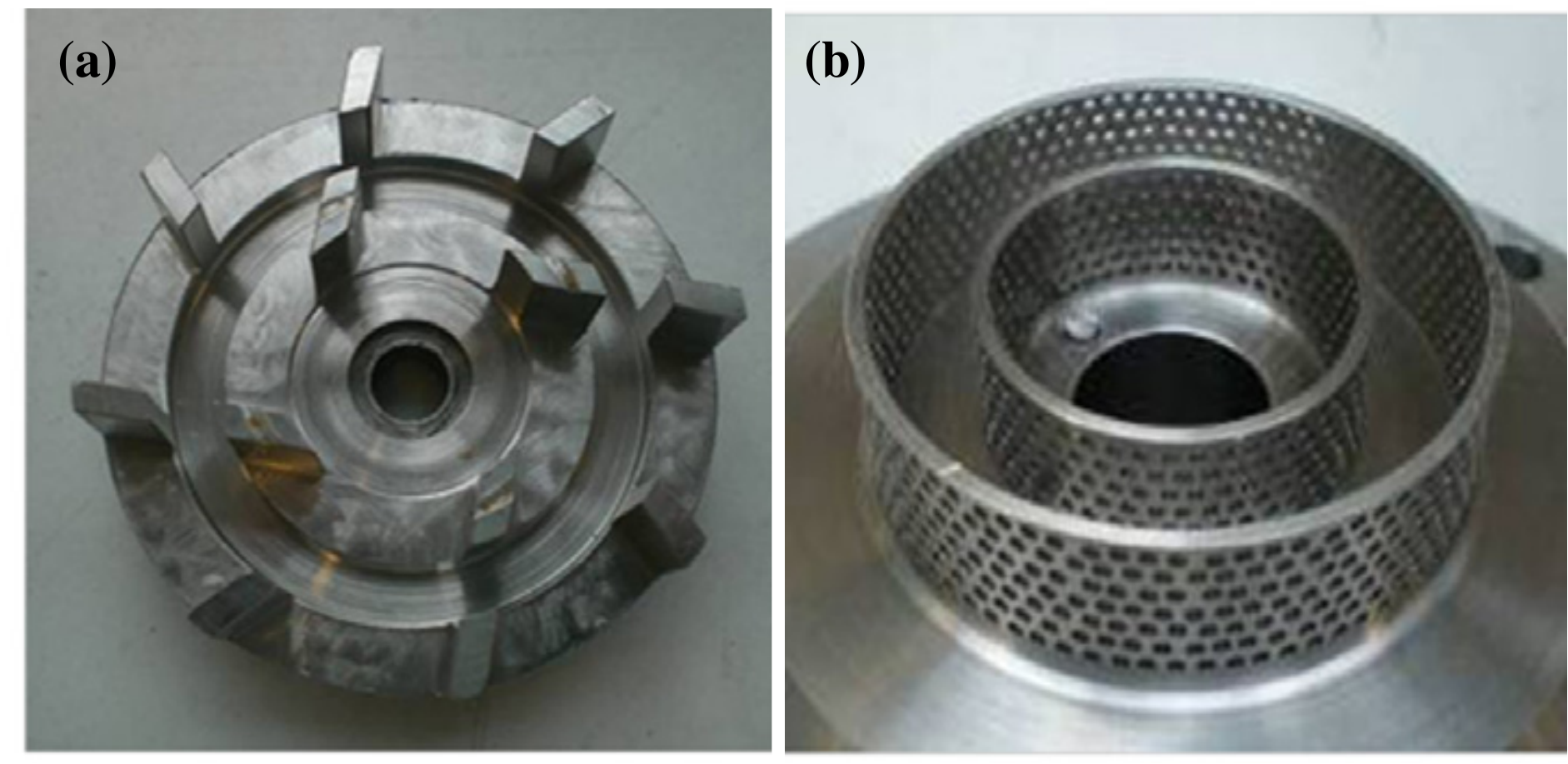}
 \caption{Silverson 150/250MS mixer, (a) Rotor (b) Stator.}
 \label{fig:Mixer_geo}
\end{figure}

The Silverson double screen 150/250MS in-line mixer has been experimentally studied in several works \cite{Cooke2008,Cooke2012,Ozcan2011,Kowalski2011,Hall2011}, measuring the power consumption and mixer performance at different operating speeds. The mixer has two rotors which rotate together within closely fitted stator screens. The rotors and stator screens of the mixer are shown in \ref{fig:Mixer_geo}. The inner screen consist of 6 rows of $50\times1.59\,\mbox{mm}$ diameter circular holes on a  triangular 2.54 mm pitch. The outer screen consist of 7 rows of $80\times1.59\,\mbox{mm}$ diameter circular holes on a  triangular 2.54 mm pitch \cite{Kowalski2011}. The inner rotor has four blades with inner diameter of 26.2 mm and an outer diameter of 38.1 mm, while the outer rotor has eight blades with an inner diameter of 49.9 mm and an outer diameter of 63.5 mm. The gap between the rotors and stator screens is 0.24 mm. The mixer usually operates over a range of speeds varying from 3000 to 12000 rpm with the fluid flowing through the device at different flow rates. 

\section{Numerical method and computational configuration} \label{sec:nmerical_code_sec}
\begin{figure}[htbp]
\centering
\includegraphics[scale=0.55]{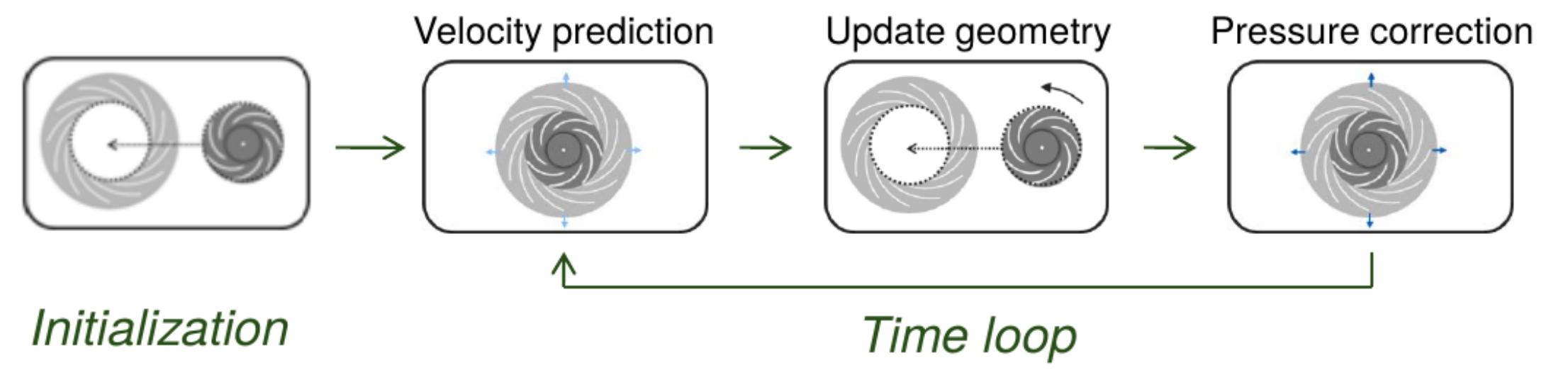}
\caption{Schematic of mesh handeling in the turbomachinery module of $Code\_Saturne$.}
 \label{fig:Turbomachinery_in_Code_Saturne}
\end{figure}

The simulations were performed using $Code\_Saturne$, an open-source CFD code developed by EDF \cite{Archambeau2004} (see http://www.code-saturne.org). $Code\_Saturne$ is an incompressible solver based on a collocated discretisation of the domain, and is able to treat structured and unstructured meshes with different cell shapes. It solves the Navier-Stokes equations with a fractional step method based on a prediction-correction algorithm
for pressure/velocity coupling (SIMPLEC), and Rhie and Chow interpolation to avoid pressure oscillations. The code uses an implicit Euler scheme for time discretisation, and a second order centred difference scheme is used for the spatial gradients. Rotating meshes are handled via a turbo-machinery module, which solves the transport equations for the initial geometry, updates the geometry and then corrects the pressure as shown in \ref{fig:Turbomachinery_in_Code_Saturne}. The code has previously been validated to many industrial and academic studies, ranging from simulations of incompressible flows (with and without rotating meshes) \cite{Ahmed2017,Jarrin2006,Jarrin2009} to low Mach number variable density reacting flows \cite{Ahmed2016,Ahmed2017a}. A number of RANS turbulence models are available in $Code\_Saturne$; the standard $k-\epsilon$ model of Jones and Launder \cite{23} with standard Log-Law wall function, the $k-\omega$ Shear Stress Transport (SST) model of Menter \cite{Menter1994} and the quasi-linear second moment closure model (SSG) of Speziale et al \cite{Speziale1991}. 

\begin{figure}[htbp]
\centering\subfigure[Model configuration in 2D]{\includegraphics[scale=0.14]{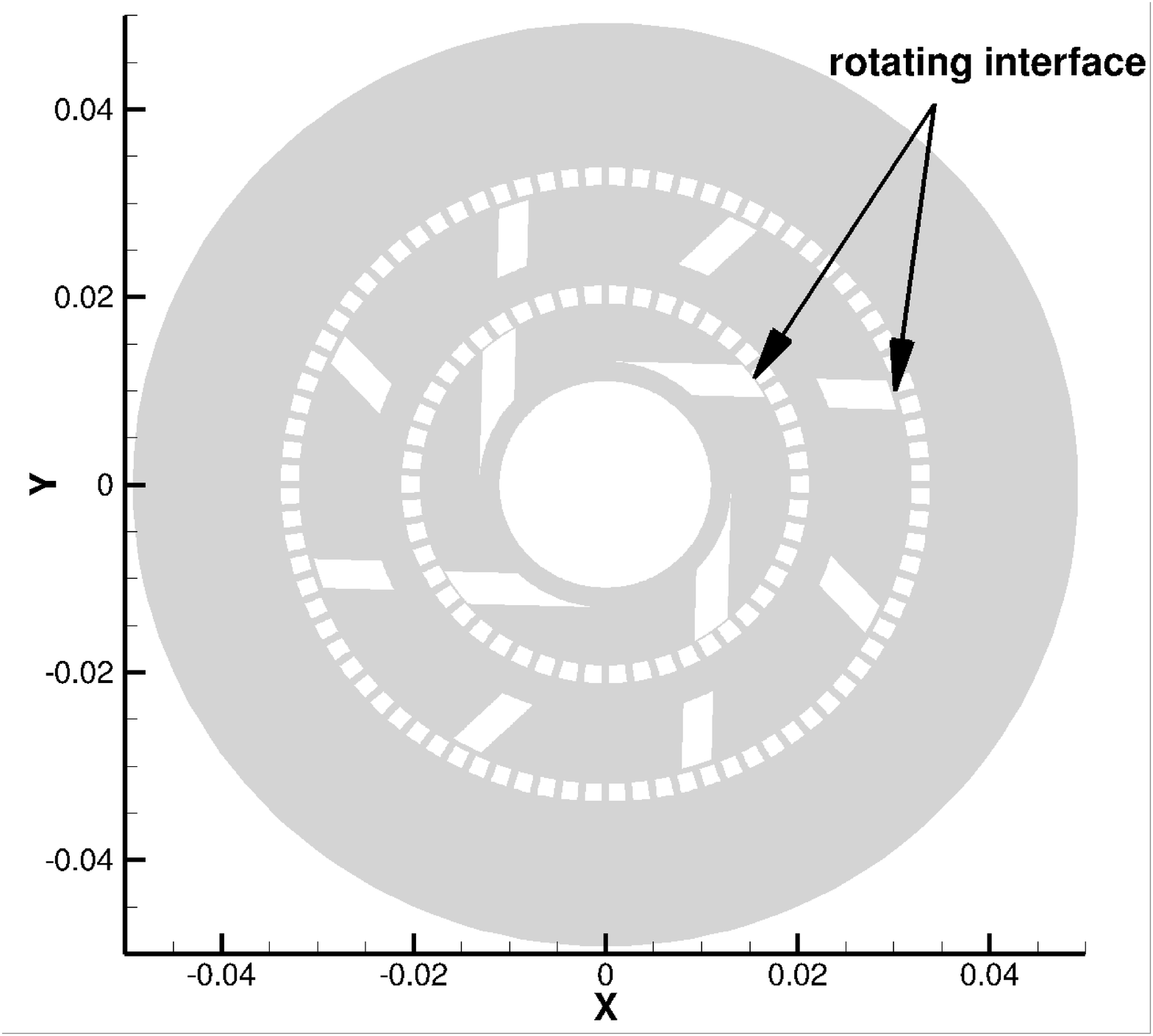}}\subfigure[Computational grid]{\includegraphics[scale=0.14]{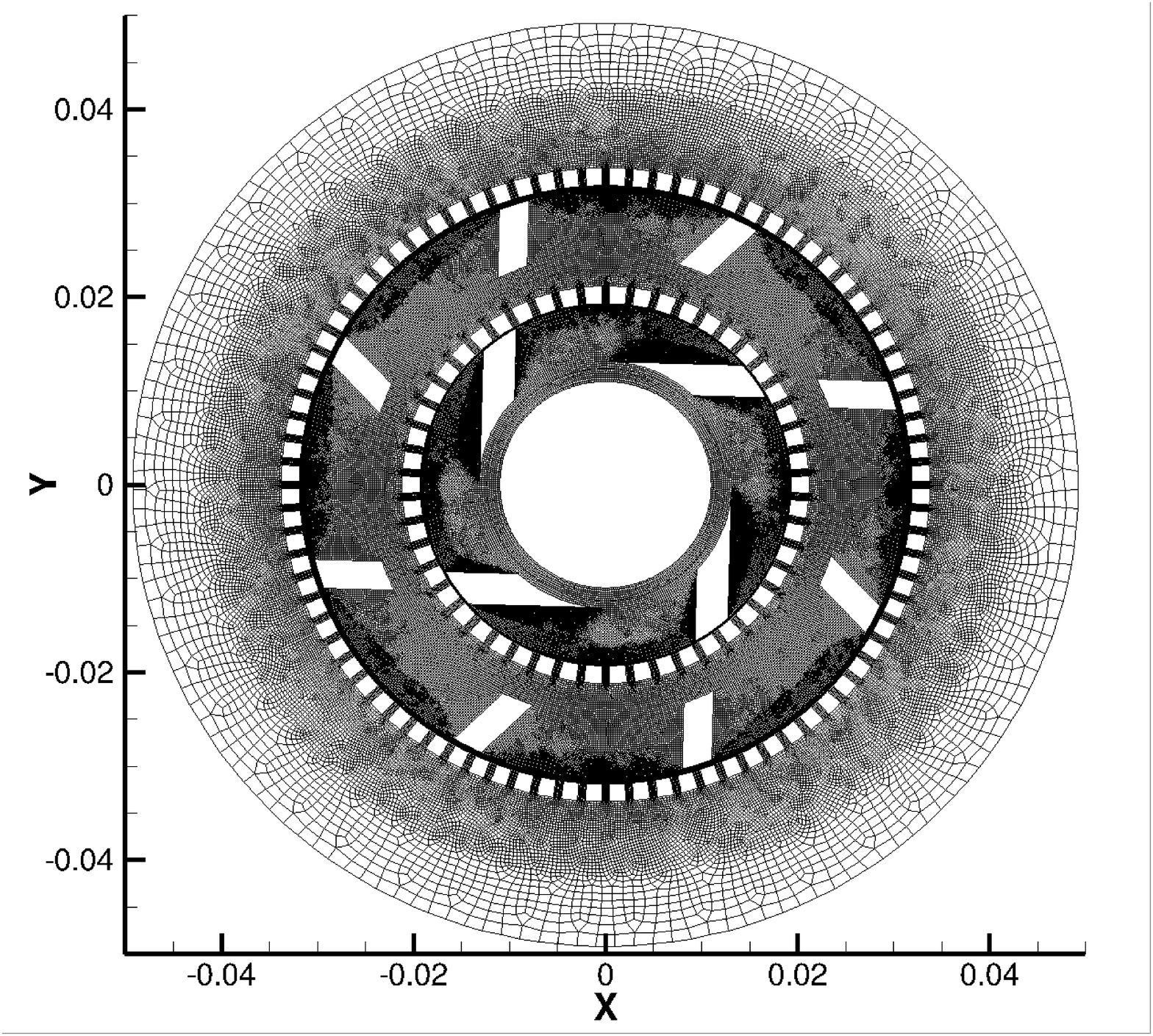}}

\centering\subfigure[Mesh near the inner rotating interface]{\includegraphics[scale=0.15]{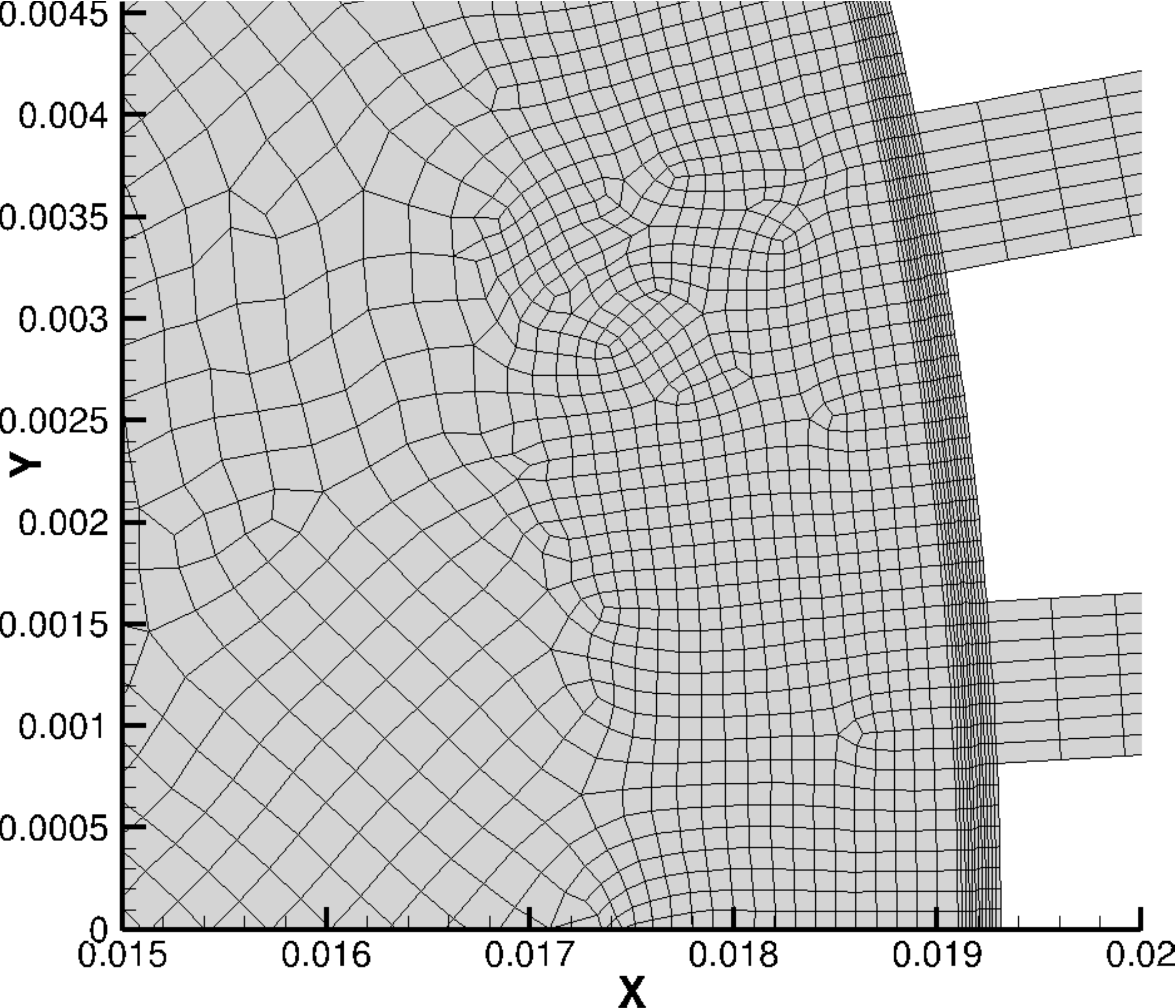}}\subfigure[Mesh near the outer rotating interface]{\includegraphics[scale=0.15]{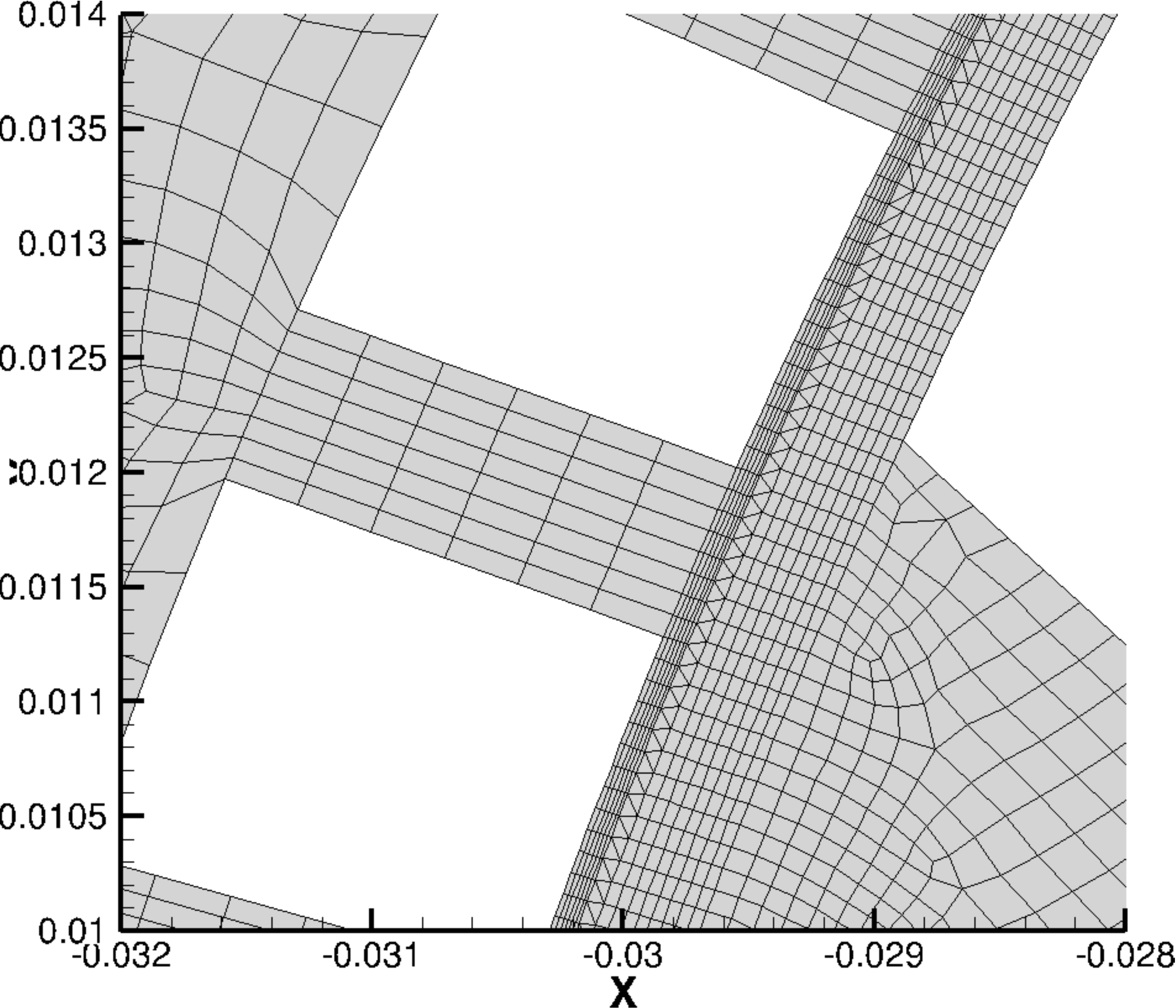}}

\caption{2D computational domain and mesh for the Silverson 150/250MS mixer \label{fig:Mesh_silverson}}
\end{figure}

A 2-D computational domain has been used in the current investigation as shown in \ref{fig:Mesh_silverson}a; this provides a comparable basis to the 2-D MRF configuration adopted in earlier studies of Jasi\'nska et al \cite{Jasinska2015,Jasinska2014}. The computational domain is meshed with $180000$ cells and shown in figure \ref{fig:Mesh_silverson}b. The mesh is refined in the regions near to the sliding interface located in the rotor-stator gaps (as shown in \ref{fig:Mesh_silverson}c and \ref{fig:Mesh_silverson}d). Grid sensitivity studies have been carried out and the grid size for mesh independent results is similar to that of Jasi\'nska et al \cite{Jasinska2015,Jasinska2014}. Standard inflow conditions on the inlet faces and pressure outlet conditions on the outlet faces are specified. A no-slip condition is applied to the velocity at the walls along with the appropriate wall treatment through standard wall functions for turbulence and zero normal gradients for scalars. Symmetry conditions are used in the transverse direction. Similar boundary conditions have been used in the earlier study of  Jasi\'nska et al \cite{Jasinska2015,Jasinska2014} for the same Silverson mixer.

\section {Turbulence models}\label{sec:Turb_models}
\subsection{Eddy viscosity models}
Eddy viscosity models rely on the turbulence viscosity hypothesis introduced by Boussinesq \cite{bouss77}:
\begin{equation}
\overline{u'_iu'_j} \propto \mu_t \frac{\partial \overline{u'_i}}{\partial x_j}.
\label{eq:bouss}
\end{equation}
The turbulent Reynolds stresses $\overline{u'_iu'_j}$ are assumed to be proportional to the mean rate of strain and the eddy viscosity $\mu_t$ is a product of length and velocity scales. The velocity scale is obtained from solving a transport equation for the turbulent kinetic energy, $k$. Depending on the choice of length scale, two of the most commonly used eddy viscosity models are the $k-\epsilon$ model where $\mu_t=\mu_t(k,\epsilon)$ \cite{23} and the $k-\omega$ SST model where $\mu_t=\mu_t(k,\omega)$ \cite{Menter1994}. Here $\epsilon$ is the turbulence energy dissipation rate and $\omega$ is the specific turbulence energy dissipation rate. The transport equations for both models are given in the appendix.

These models are simple and effective in terms of computational cost, but have some predictive failings, including where flows exhibit strong turbulent stress anisotropy. Flows with strong rotation and curvature effects and flows with complex strain fields (such as those found in the Silverson mixer) historically have been challenging to capture via eddy viscosity models \cite{22}. The problem arises from trying to characterise the complex stress state embodied in $\overline{u'_iu'_j}$ via \ref{eq:bouss}, even though the turbulent kinetic energy $k=\overline{u'_iu'_i}$ is computed to a reasonable accuracy \cite{22}. To rectify this shortcoming second moment closure models are needed.

\subsection{Second moment closure models}
Several major drawbacks of the eddy viscosity models are overcome by second moment closures or Reynolds stress transport models (RSM). In these models, transport equations for the six independent components of the Reynolds stress tensor and an additional equation for turbulent dissipation $\epsilon$ are solved. These models are able to account for anisotropies in the Reynolds stress field without further modelling. One of the most widely used second moment closure models is the quasi-linear closure model of Speziale et al \cite{Speziale1991}, commonly known as the SSG model. The details of the equations solved in these calculations are given in the appendix.

Second moment closures generally lead to significant improvements in the prediction of mean flow properties and of the Reynolds stresses for simple and complex flows (i.e. wall jets, asymmetric channels and curved flows) \cite{22, Launder2011}. One of the major drawbacks of second moment closure models is their associated computational cost;  these models have consequently not been widely used in industrial flows. Additionally, the models' sophistication can lead to numerical convergence problems due to the coupling of the mean velocity and turbulent stress fields through source terms \cite{22}; their use typically requires users with greater degree of technical CFD awareness. 

\section{Results and discussion}\label{sec:results}
The results from using different turbulence models and different algorithms for handling the rotation of the mixer are reported in this section. The simulations are compared with the experimental results of Kowalski et al \cite{Kowalski2011} for power consumption and Cooke et al \cite{Cooke2012} for power number at different Reynolds numbers.

\subsection{Comparison between different solution methods}

\begin{figure}[htbp]
\centering\includegraphics[scale=0.1]{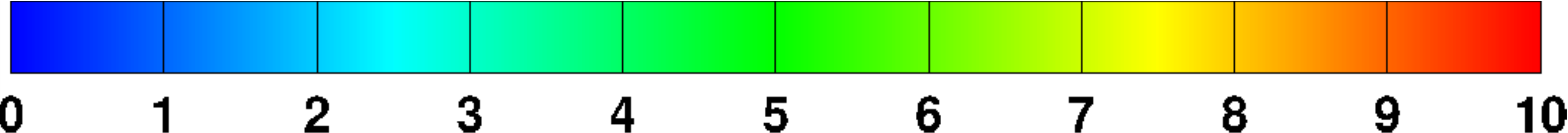}\\
\vspace{0.01cm}
\centering\subfigure[Multiple reference frame]{\includegraphics[scale=0.15]{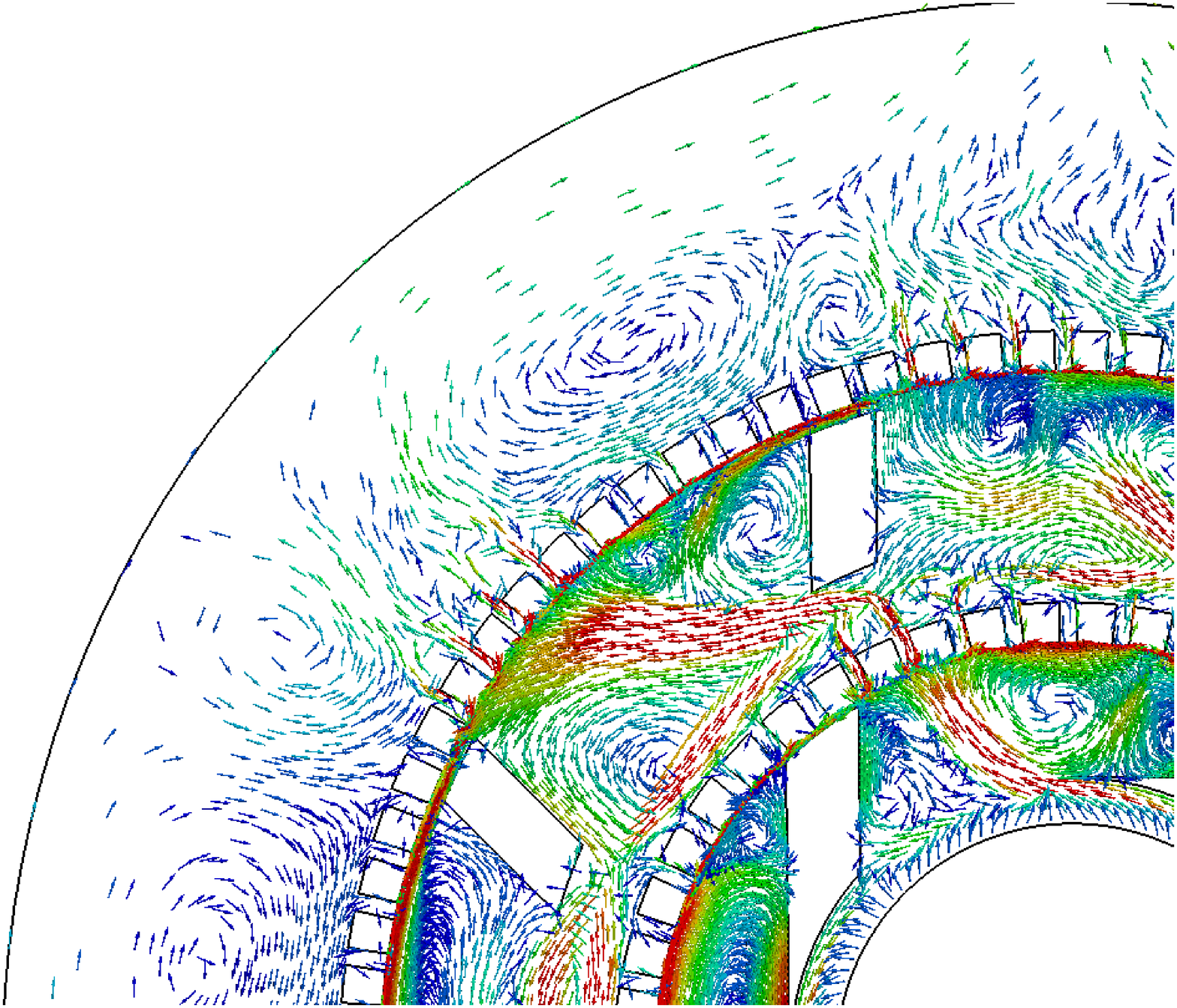}}
\vspace{0.01cm}
\centering\subfigure[Sliding mesh algorithm]{\includegraphics[scale=0.15]{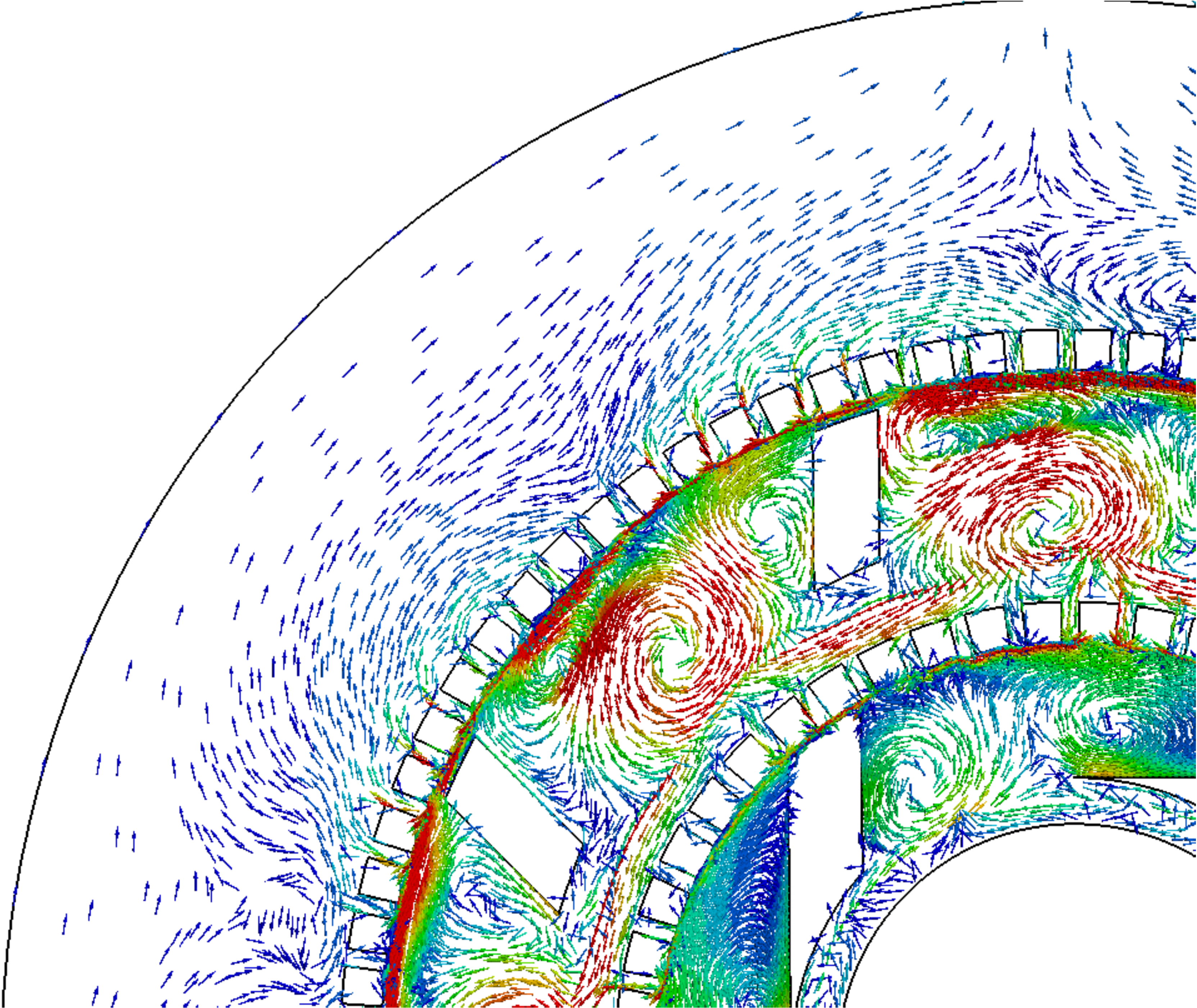}}
\caption{Relative velocity vector (m/s) prediction by using the multiple reference frame and the sliding mesh algorithm. Mixer is rotating at 6000 rpm. \label{fig:MRF_VS_Rotating_mesh_vec}}
\end{figure}

Two different methods to account for the rotation of mixer are first compared. \ref{fig:MRF_VS_Rotating_mesh_vec} shows the relative velocity predictions produced using the MRF and sliding mesh methods. It can be seen that the MRF method leads to the formation of an anomalous jet between the inlet and the outer screen (\ref{fig:MRF_VS_Rotating_mesh_vec}a). This leads to the formation of large recirculation zones between the outer screen and the mixer wall as shown in \ref{fig:MRF_VS_Rotating_mesh_vec}a. These jets and recirculation regions have been reported in the earlier simulations of Jasi\'nska et al \cite{Jasinska2015}. These jets are anomalous because they oppose the direction of rotation and lead to the formation of a recirculation zone on the pressure side of the mixer blade. They form because the fixed rotor-stator configuration in the presence of imposed Coriolis body forces (via MRF) provide a spurious curvilinear leak path for the fluid. The observed recirculation is an equally spurious response to this driver. 

The rotating mesh algorithm (\ref{fig:MRF_VS_Rotating_mesh_vec}b) overcomes this problem by physically accounting for periodic passing of the rotor blades over the holes of the stator; the size of the recirculation zones between the outer screen and the mixer wall is hence substantially reduced. \ref{fig:MRF_VS_Rotating_mesh_vec}b shows recirculation zones forming at the back of the mixer blades. This leads to a pressure drop across the blade which in turn leads to more force on the blade resulting in high power consumption. 

\begin{figure}[htbp]
\centering\subfigure[Initial distribution of concentration]{\includegraphics[scale=0.1]{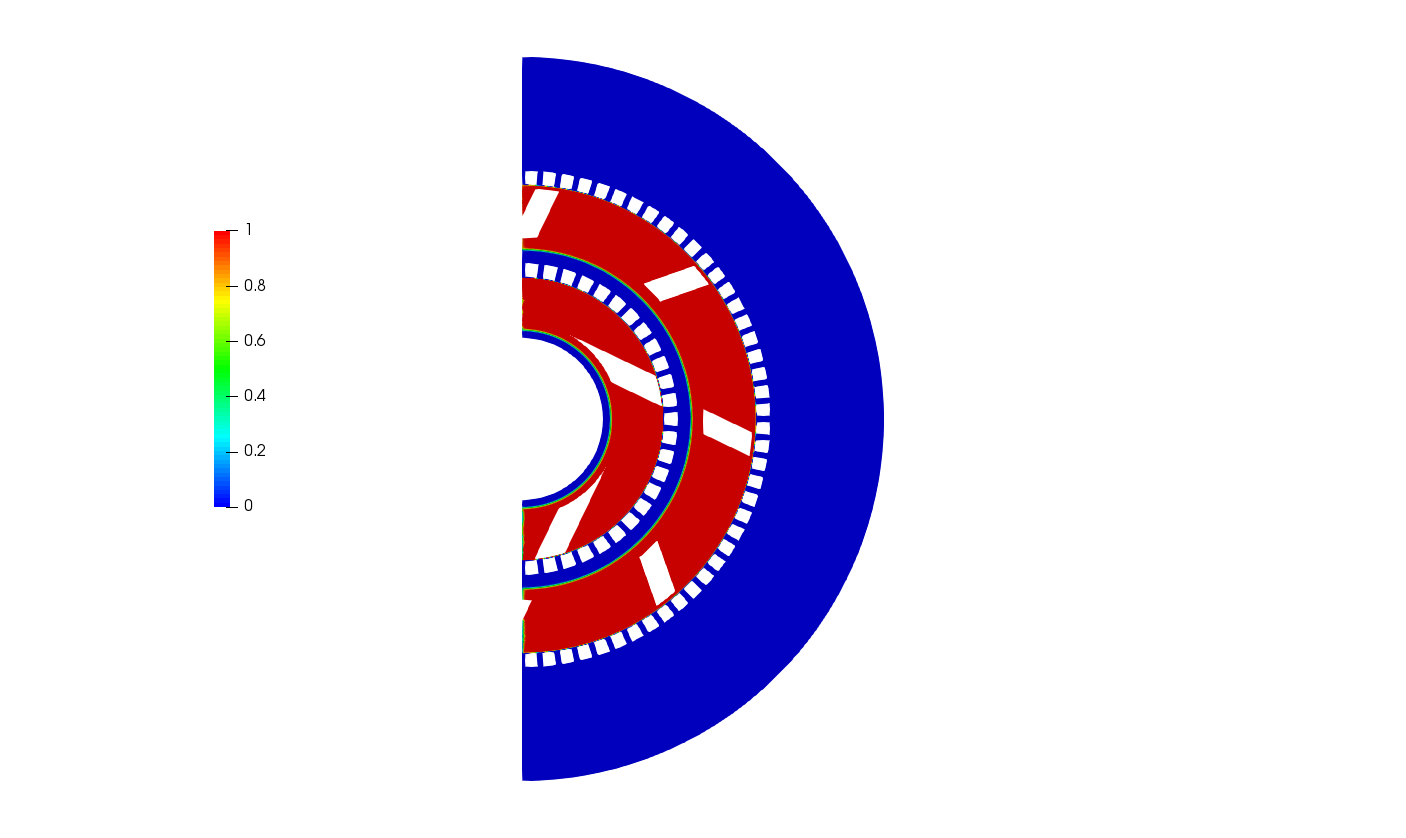}} \label{fig:Scalar_m_t0}\\
\centering\subfigure[MRF method]{\includegraphics[scale=0.1]{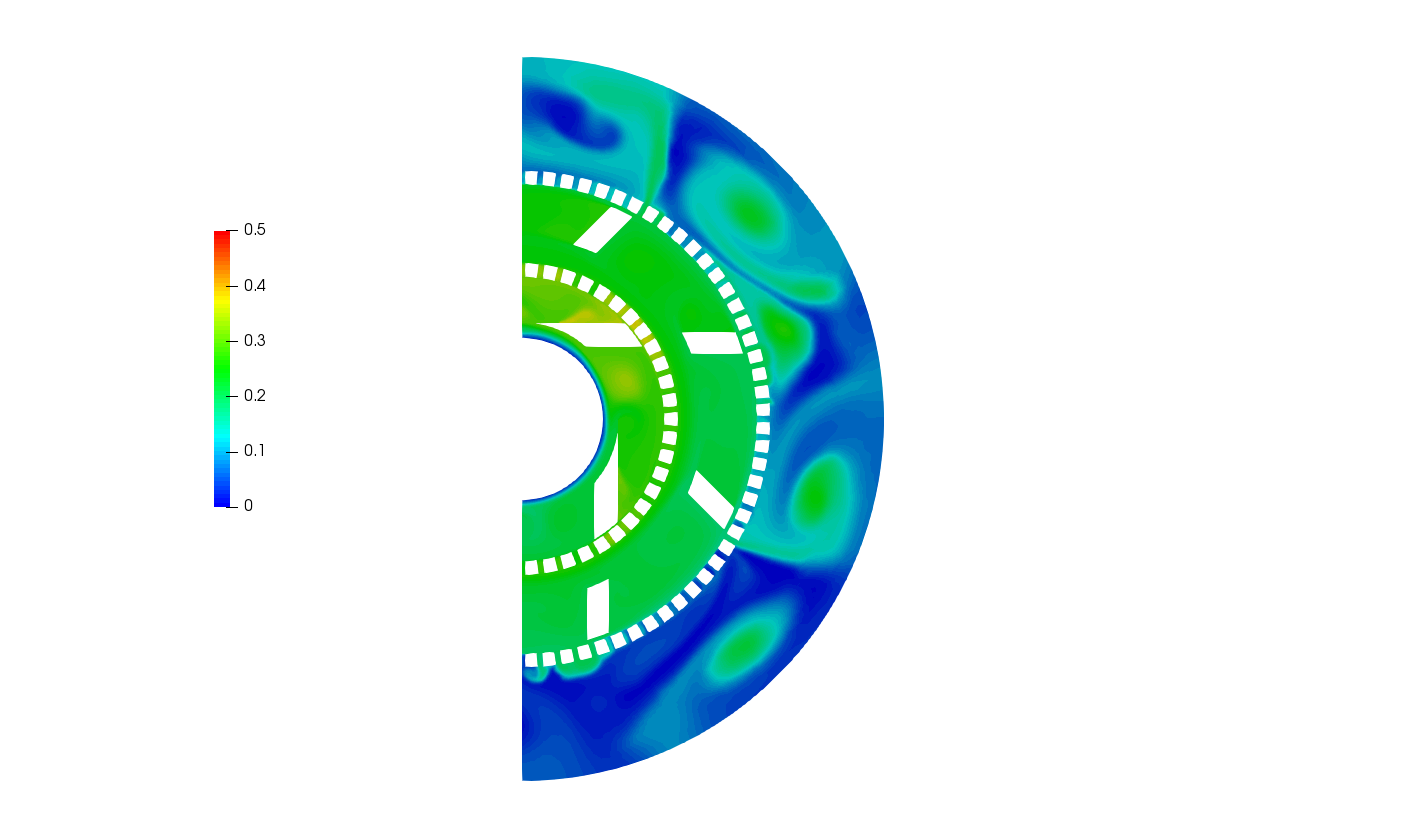}}\label{fig:Scalar_mrf} 
\centering\subfigure[Sliding mesh method]{\includegraphics[scale=0.1]{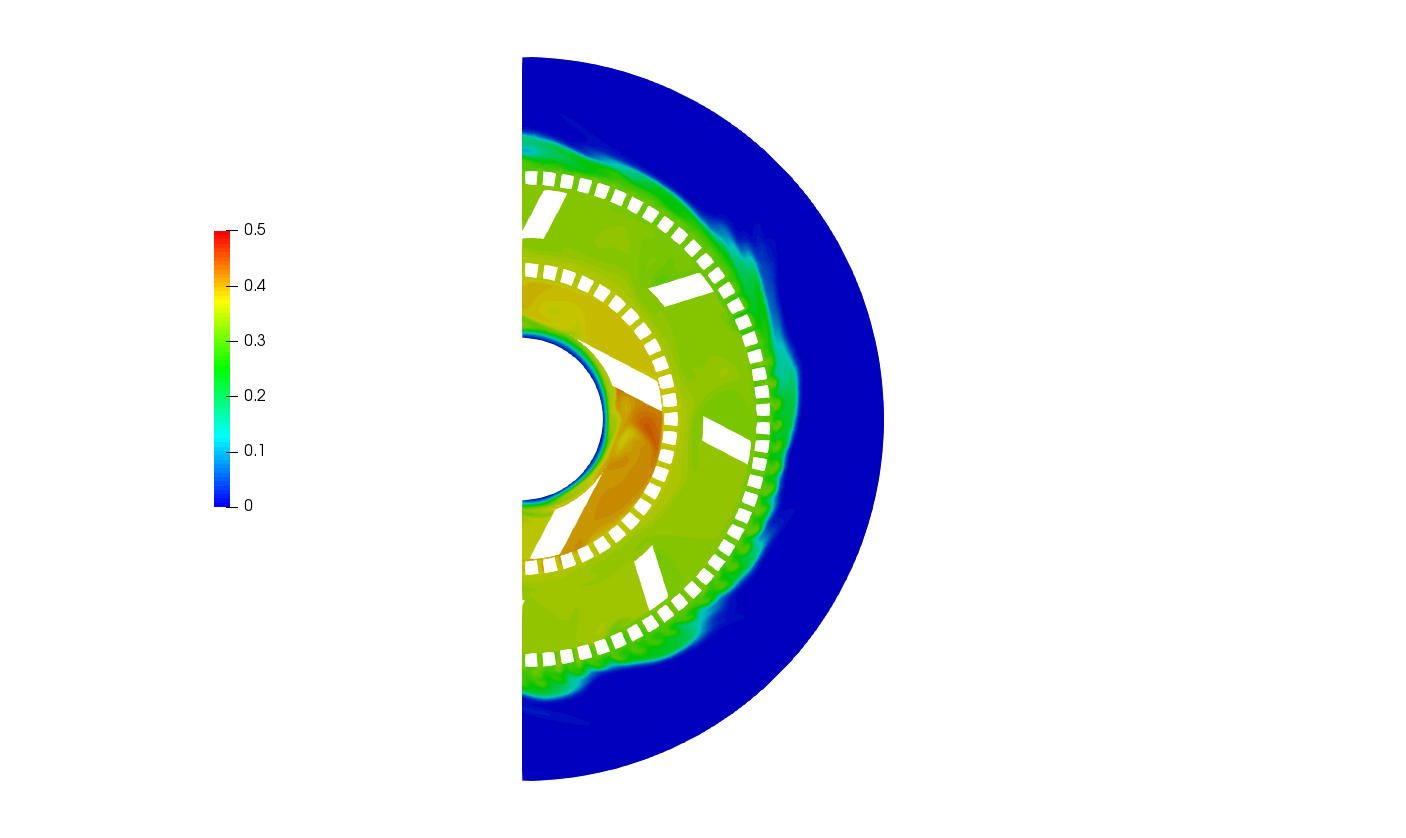}}\label{fig:Scalar_sm}
\caption{Distribution of scalar concentration after 10 revolutions using different solution methods with SSG model. Mixer is rotating at 6000 rpm with zero inflow.} \label{fig:Scalar_m}
\end{figure}

The influence of the solution method on the prediction of mixing within the mixer is shown in \ref{fig:Scalar_m}. These results have been obtained using the SSG model. The MRF method (\ref{fig:Scalar_m}b) is shown to overestimate mixing by predicting a more homogenous distribution of the scalar within the rotor swept volume compared to the sliding mesh method (\ref{fig:Scalar_m}c). The influence of the unphysical preferential leak paths predicted by the MRF can also be observed beyond the outer screen. 

\begin{figure}[htbp]
\centering
\includegraphics[scale=0.45]{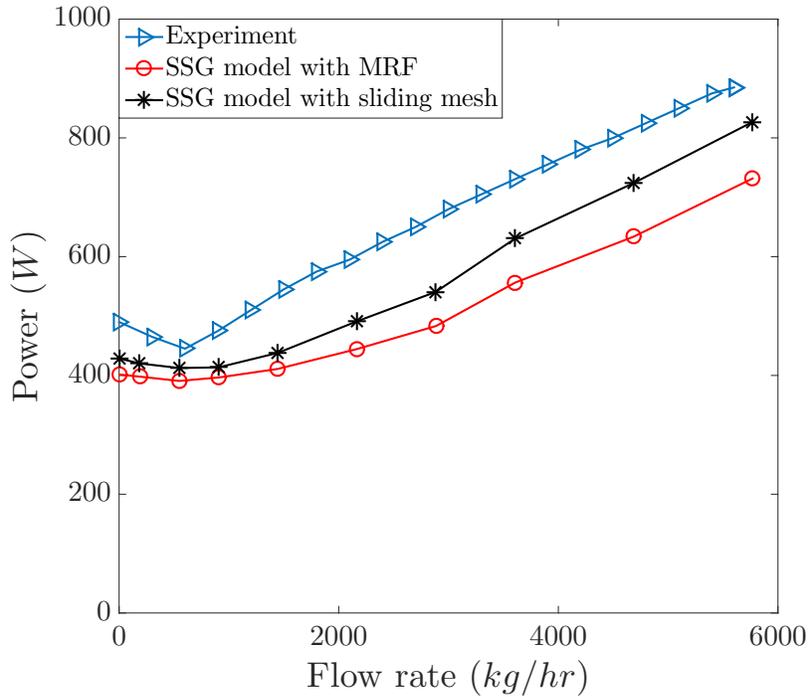}
 \caption{Power curve for Silverson 150/250 mixer using MRF and sliding mesh solution methods.}
 \label{fig:MRF_vs_sliding_mesh}
\end{figure}

\ref{fig:MRF_vs_sliding_mesh} shows the power prediction from the MRF method and the rotating mesh algorithm each using the SSG turbulence model. The predicted power from the simulations is calculated as \cite{Cooke2012}: 
 \begin{equation}
P=2\pi NM,\label{eq:Power}
\end{equation}
where $N$ is the rotation speed and $M$ is the torque calculated from the simulations.  
Both methods predict an increase in power as the flow rate increases which is consistent with the experimental results of Kowalski et al \cite{Kowalski2011}. There is a small discrepancy in the predicted trend at low flow rates ($Q<500$ kg/hr) where the power in the experiment decreases due to a rapid drop in the mixer pumping efficiency \cite{Kowalski2011}. Both methods in the simulation fail to predict this trend accurately due to the use of a steady flow rate assumption used at the inlet boundary.  It can be seen in \ref{fig:MRF_vs_sliding_mesh} that there is a very small difference between the two methods at lower flow rates and that this difference increases as the flow rate is increased. The results show that the rotating mesh algorithm improves the power prediction by $5\%$ at lower flow rates and by $12.5\%$ at the higher flow rates when compared to the MRF method. Overall it may be seen that the rotating mesh algorithm leads to power predictions that are closer to experimental results. Both CFD methods however underpredict power compared to experiments and this could be related to using a 2-D representation of a 3-D flow field. 

\subsection{Turbulence model effects}
\subsubsection{Flow and mixing characterisation}
\begin{figure}[htbp]
\centering\subfigure[$k-\omega$ SST at Q=0 kg/hr]{\includegraphics[scale=0.13]{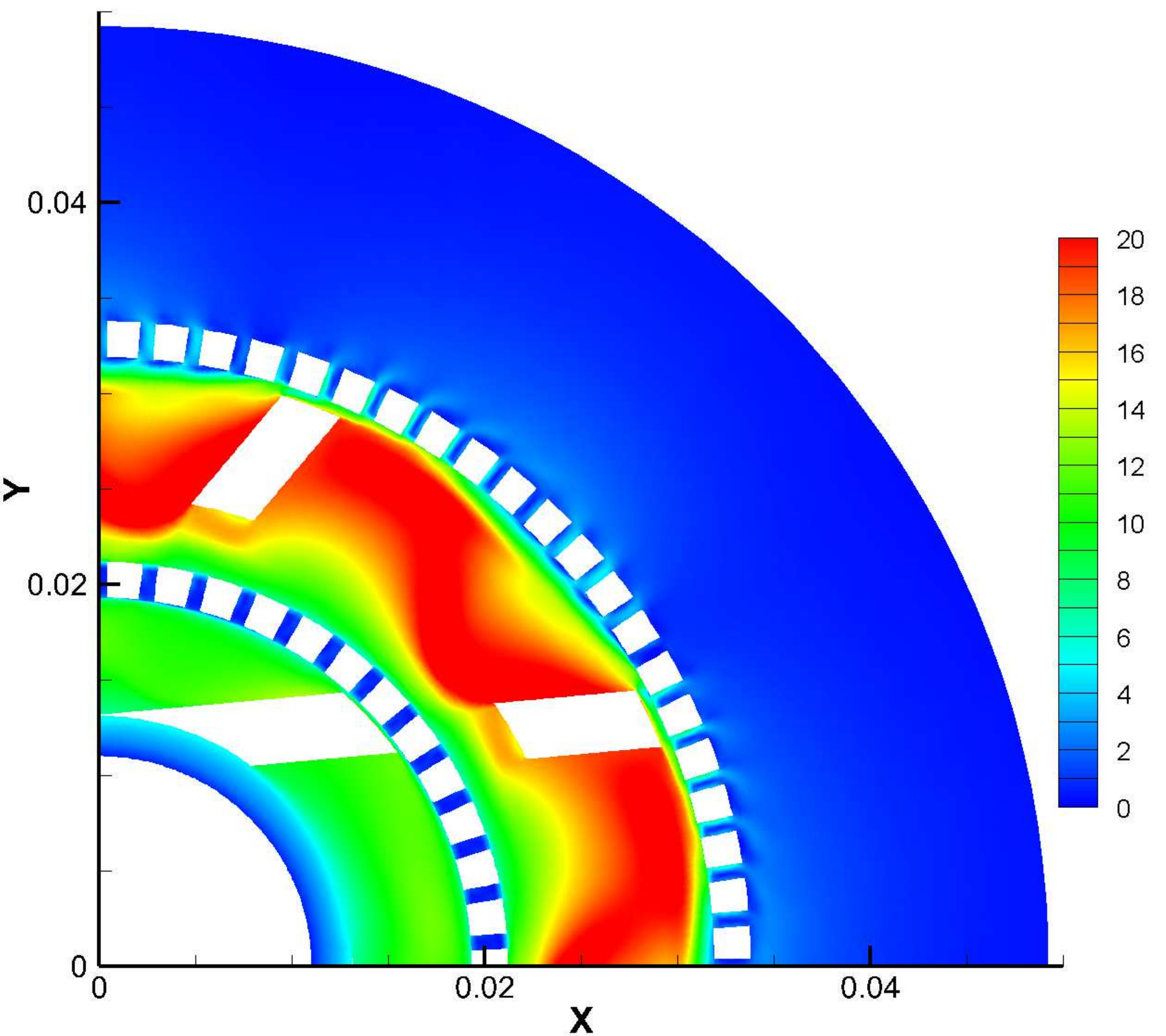}}\subfigure[SSG at Q=0 kg/hr]{\includegraphics[scale=0.13]{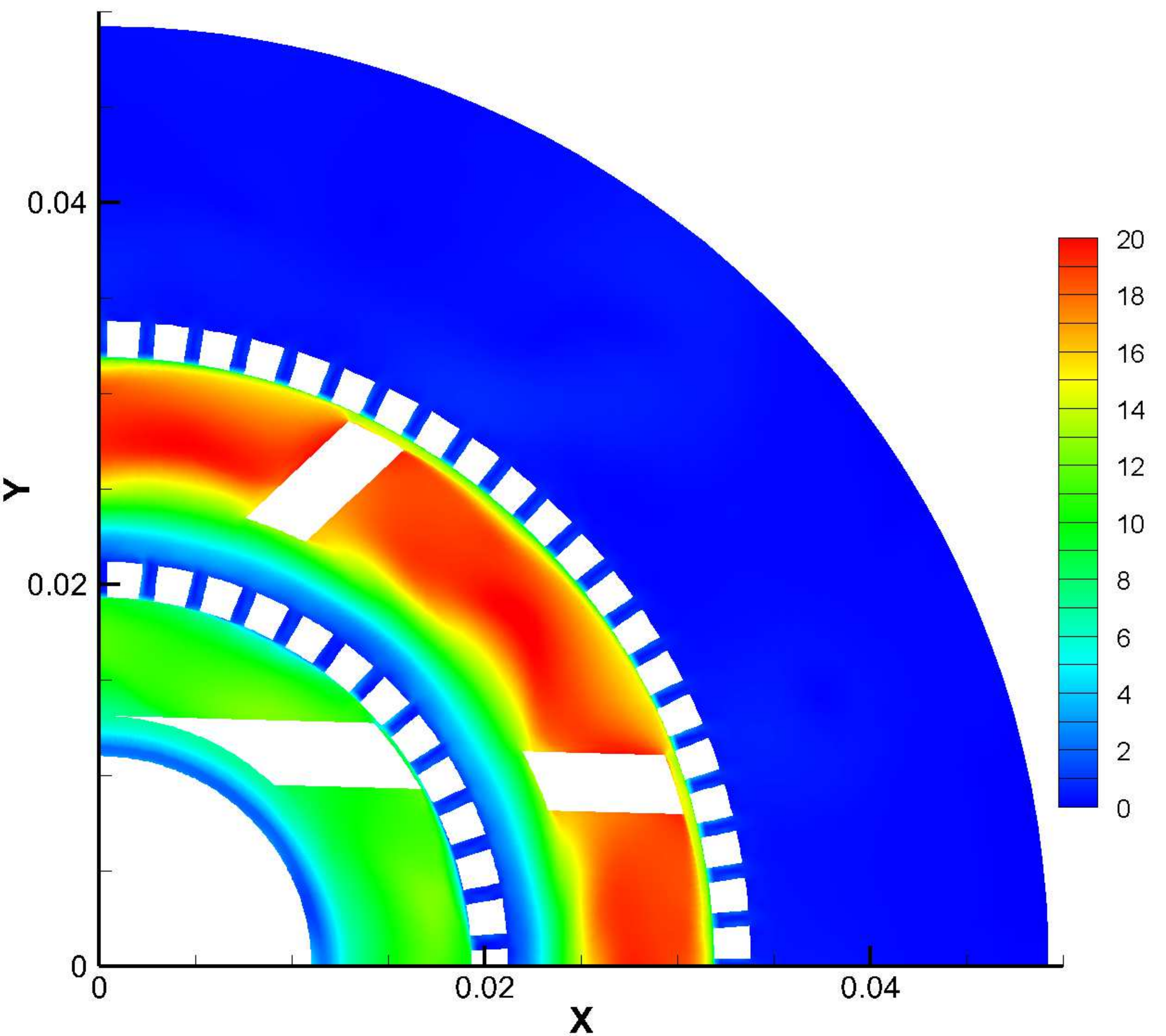}}

\centering\subfigure[$k-\omega$ SST at Q=500 kg/hr]{\includegraphics[scale=0.13]{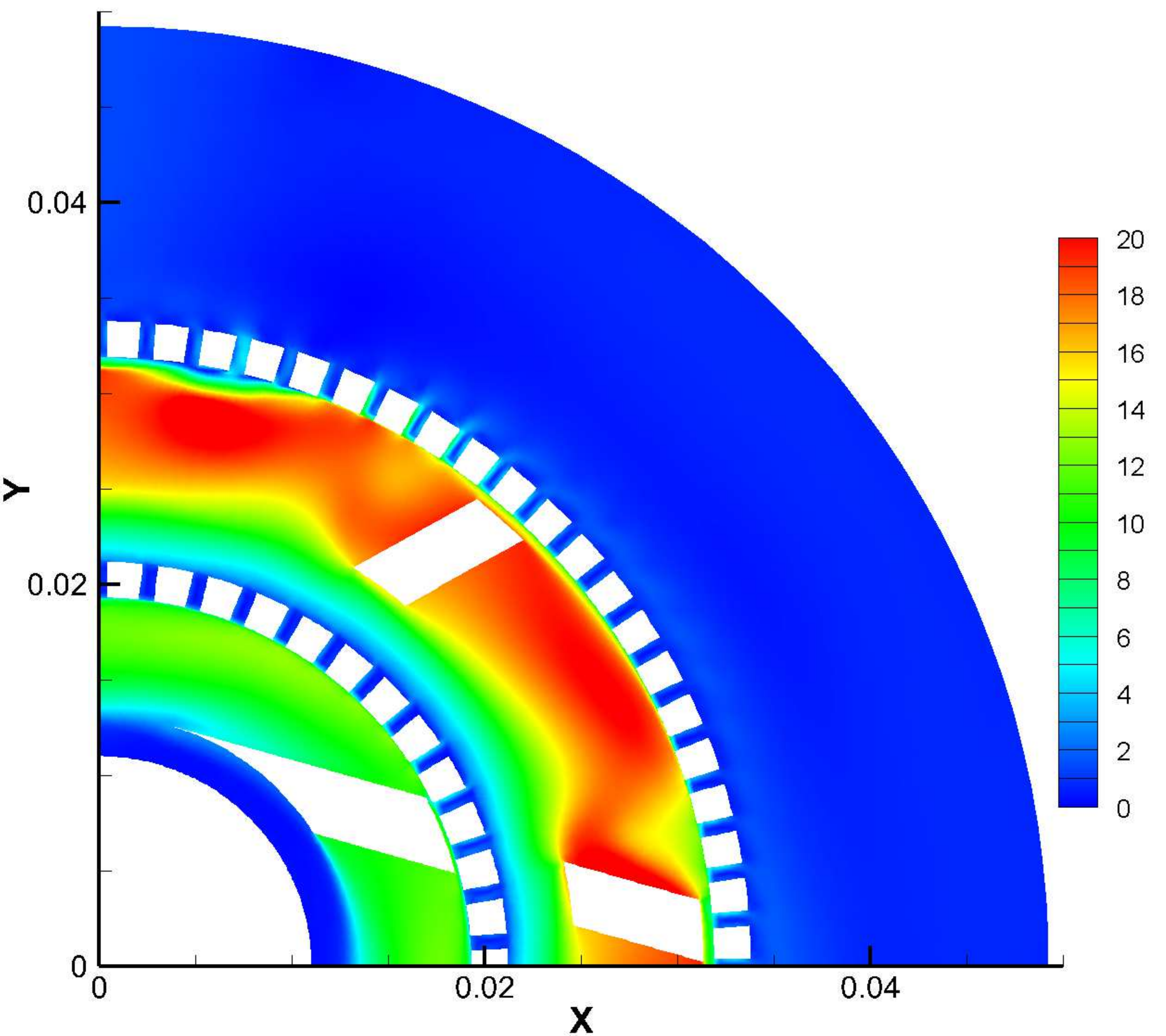}}\subfigure[SSG at Q=500 kg/hr]{\includegraphics[scale=0.13]{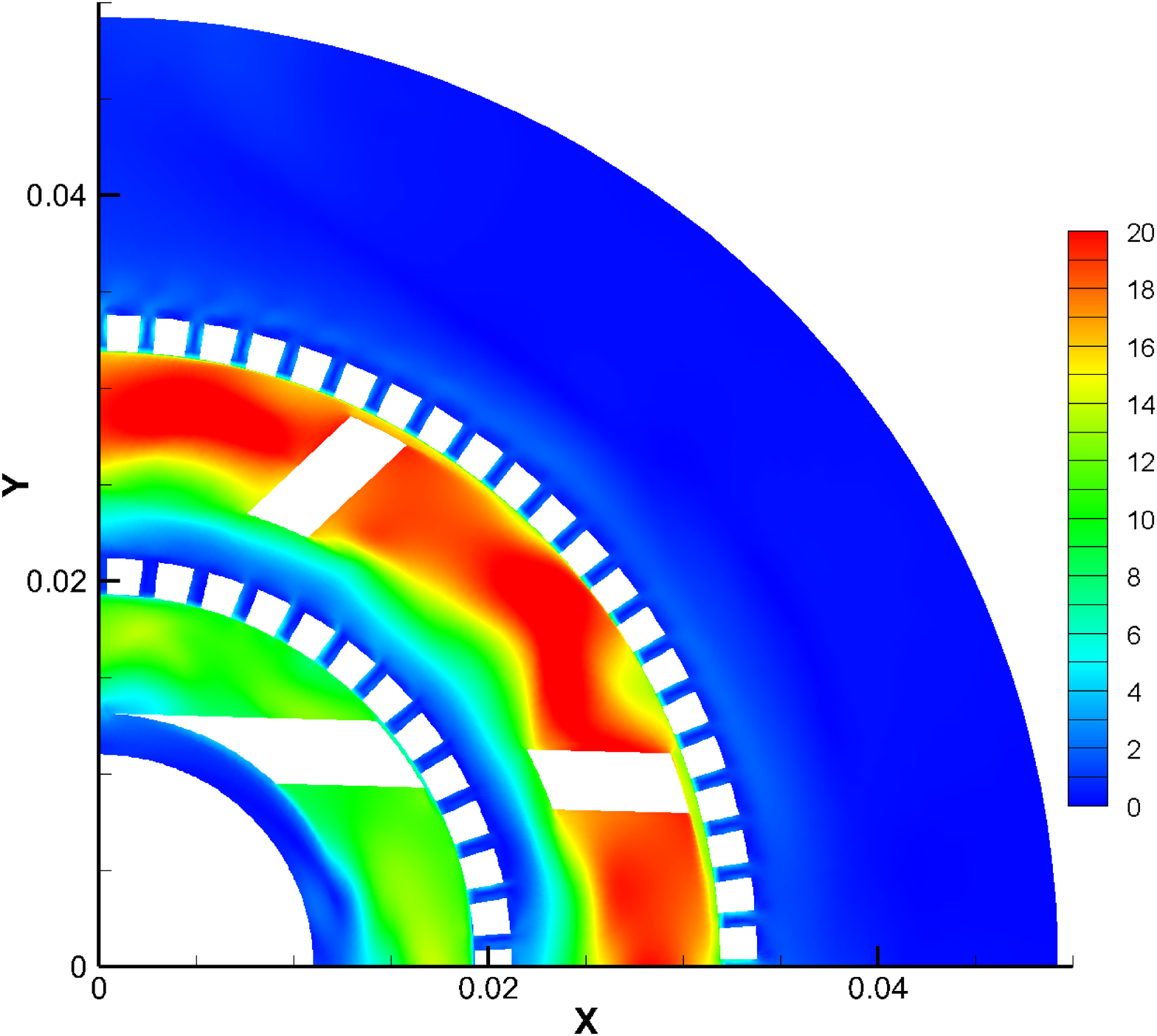}}

\centering\subfigure[$k-\omega$ SST at Q=3000 kg/hr]{\includegraphics[scale=0.13]{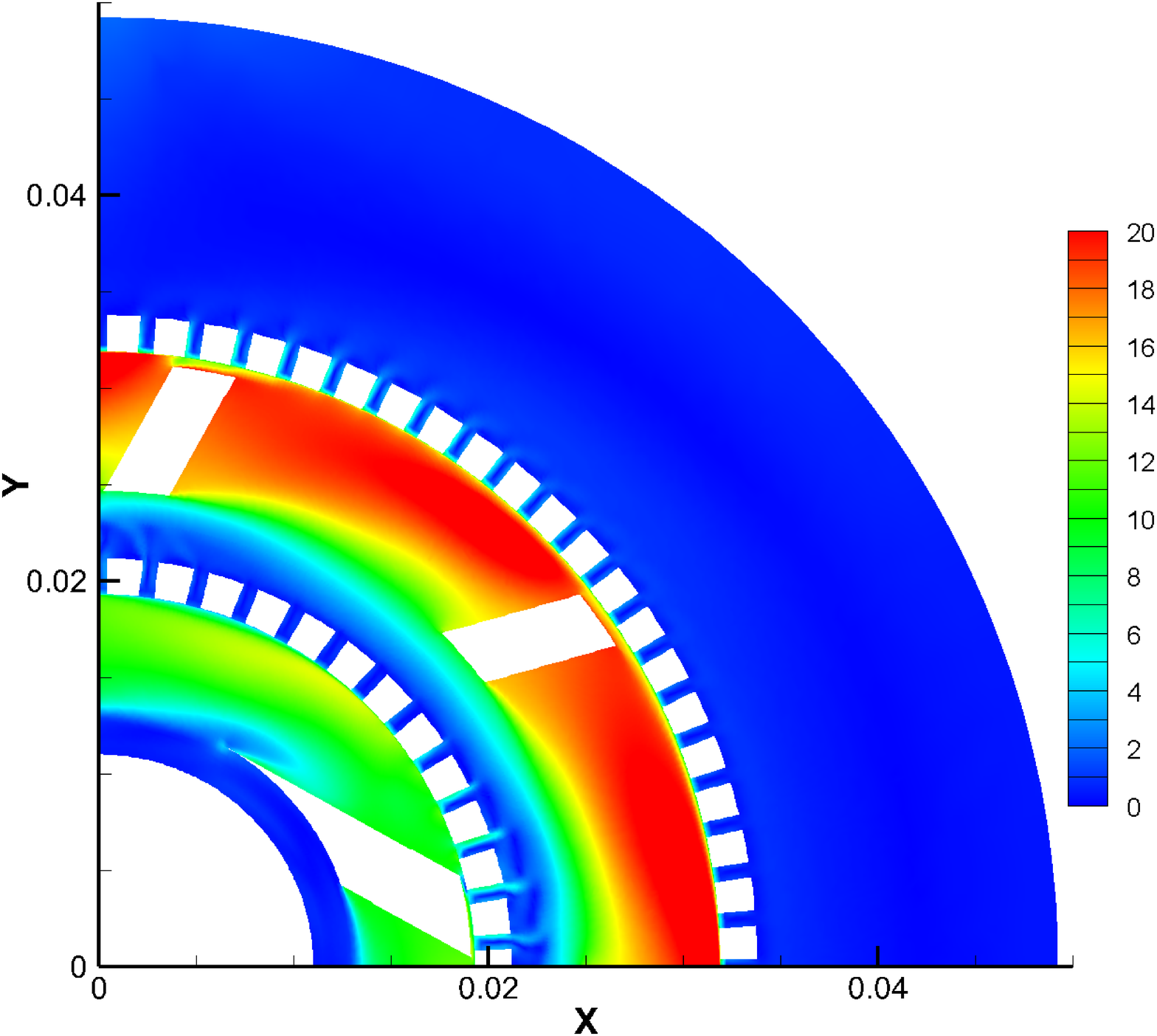}}\subfigure[SSG at Q=3000 kg/hr]{\includegraphics[scale=0.13]{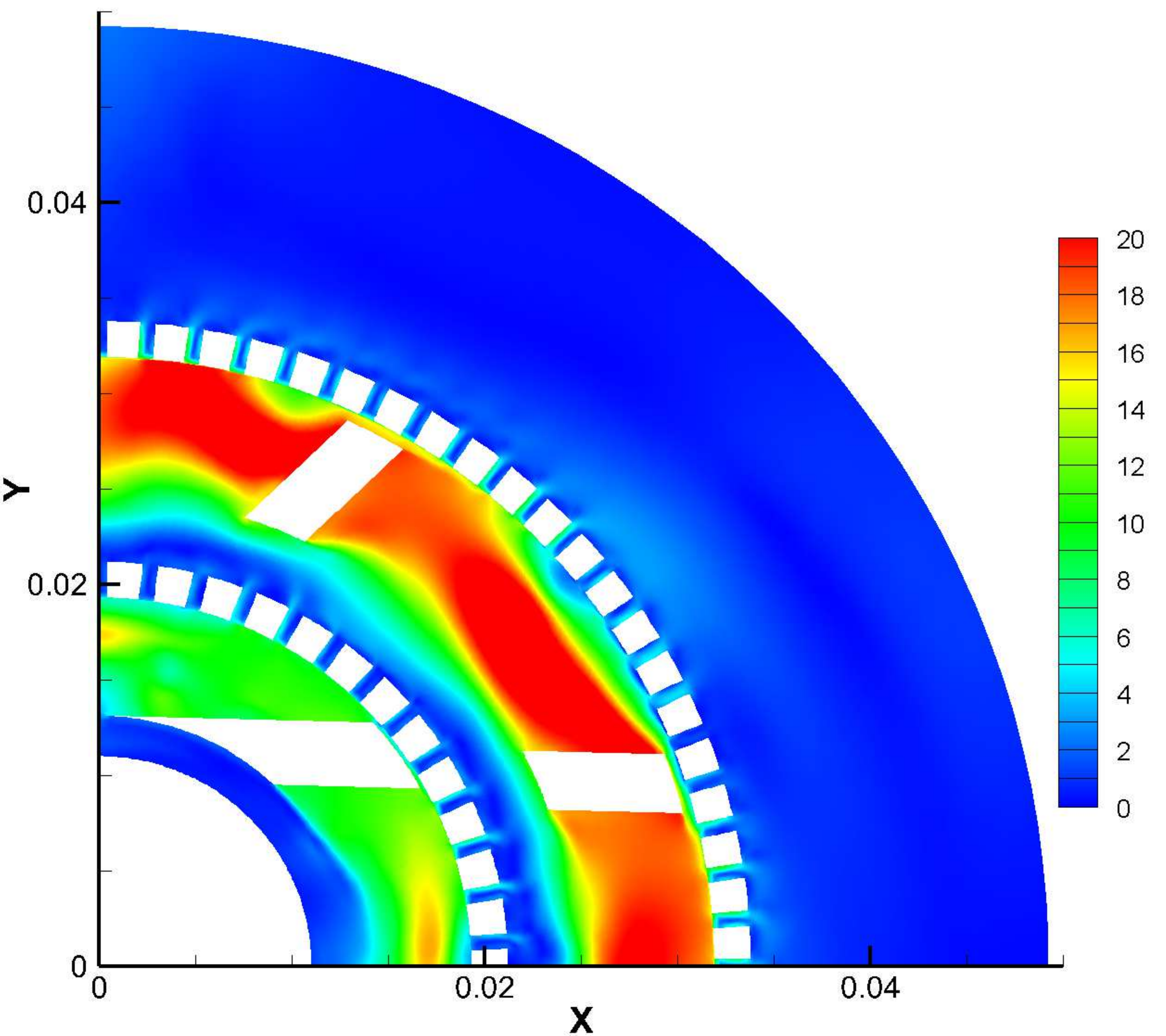}}

\caption{ Velocity predictions from $k-\omega$ SST model and SSG model.\label{fig:Vel_SST_vs_SSG}}
\end{figure}

\begin{figure}[htbp]
\centering\subfigure[$k-\omega$ SST at Q=0 kg/hr]{\includegraphics[scale=0.13]{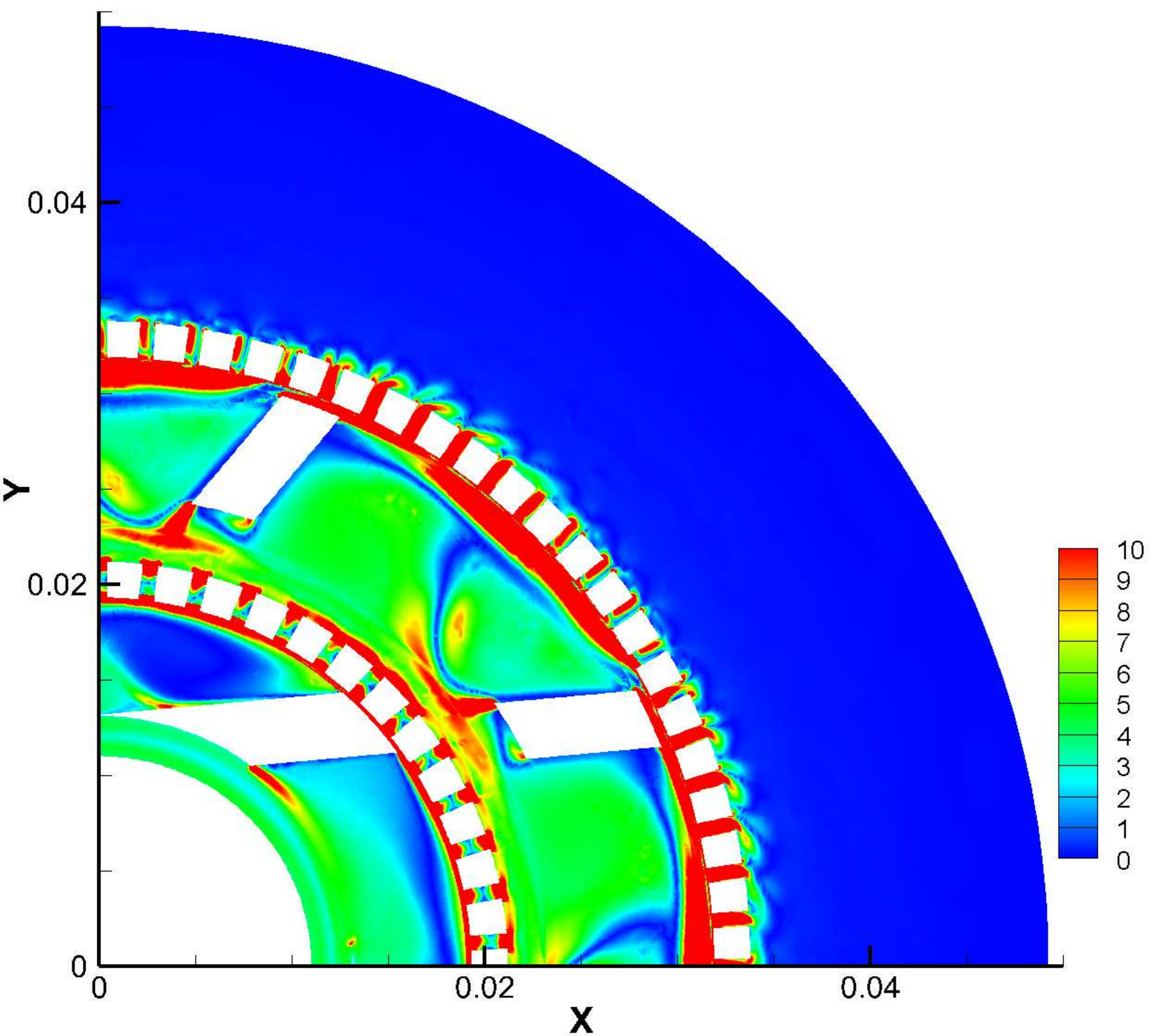}}\subfigure[SSG at Q=0 kg/hr]{\includegraphics[scale=0.13]{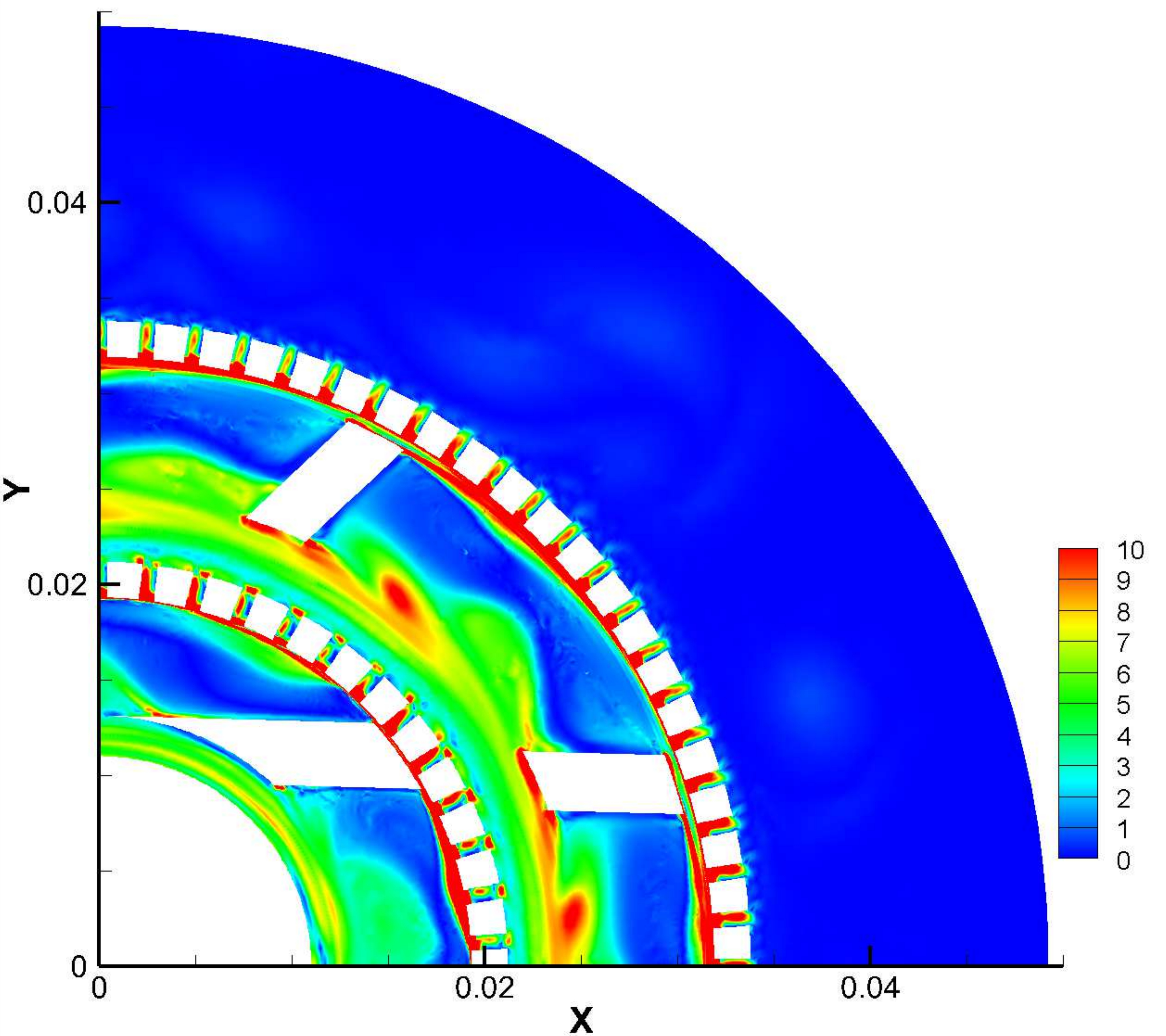}}

\centering\subfigure[$k-\omega$ SST at Q=500 kg/hr]{\includegraphics[scale=0.13]{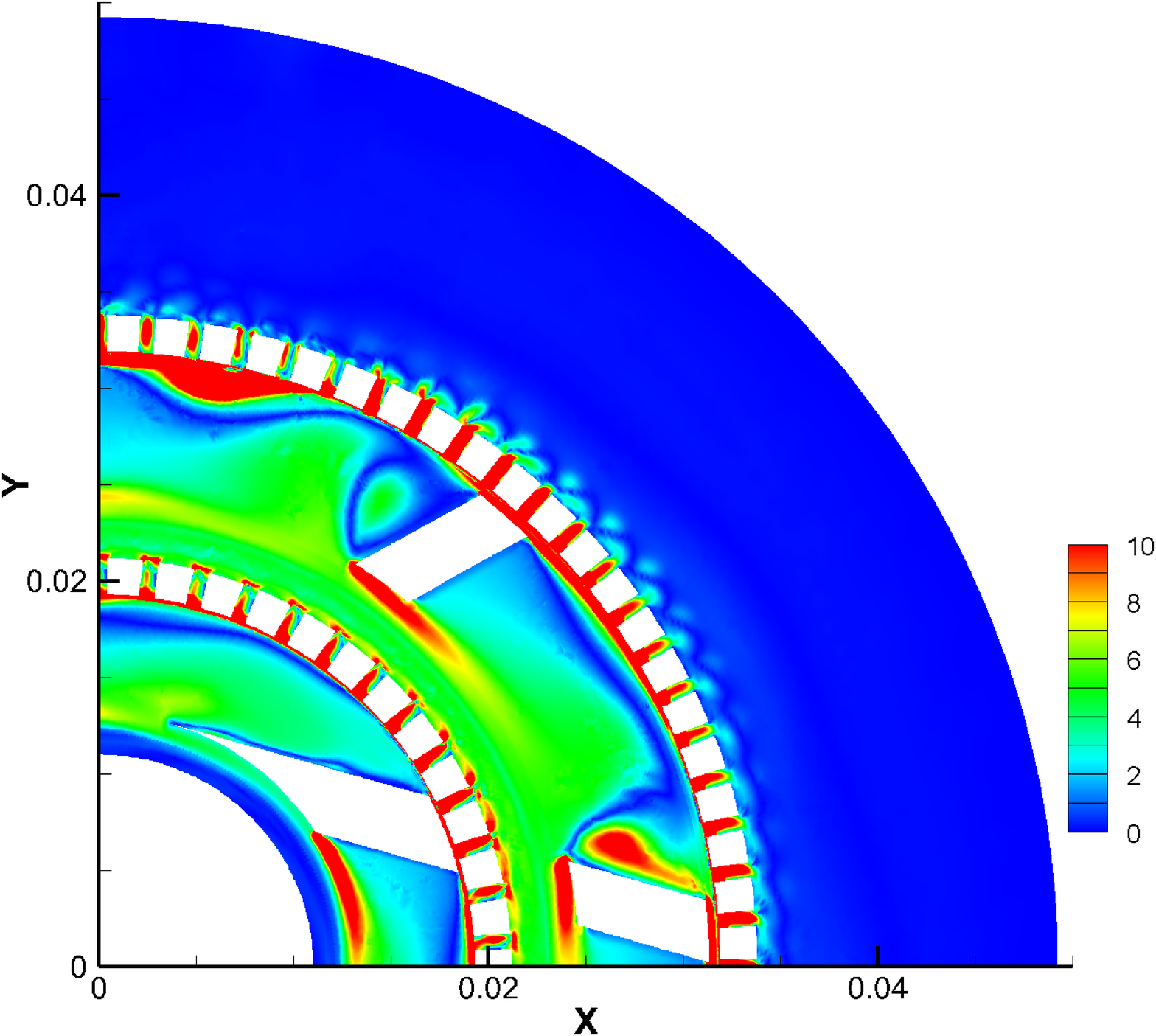}}\subfigure[SSG at Q=500 kg/hr]{\includegraphics[scale=0.13]{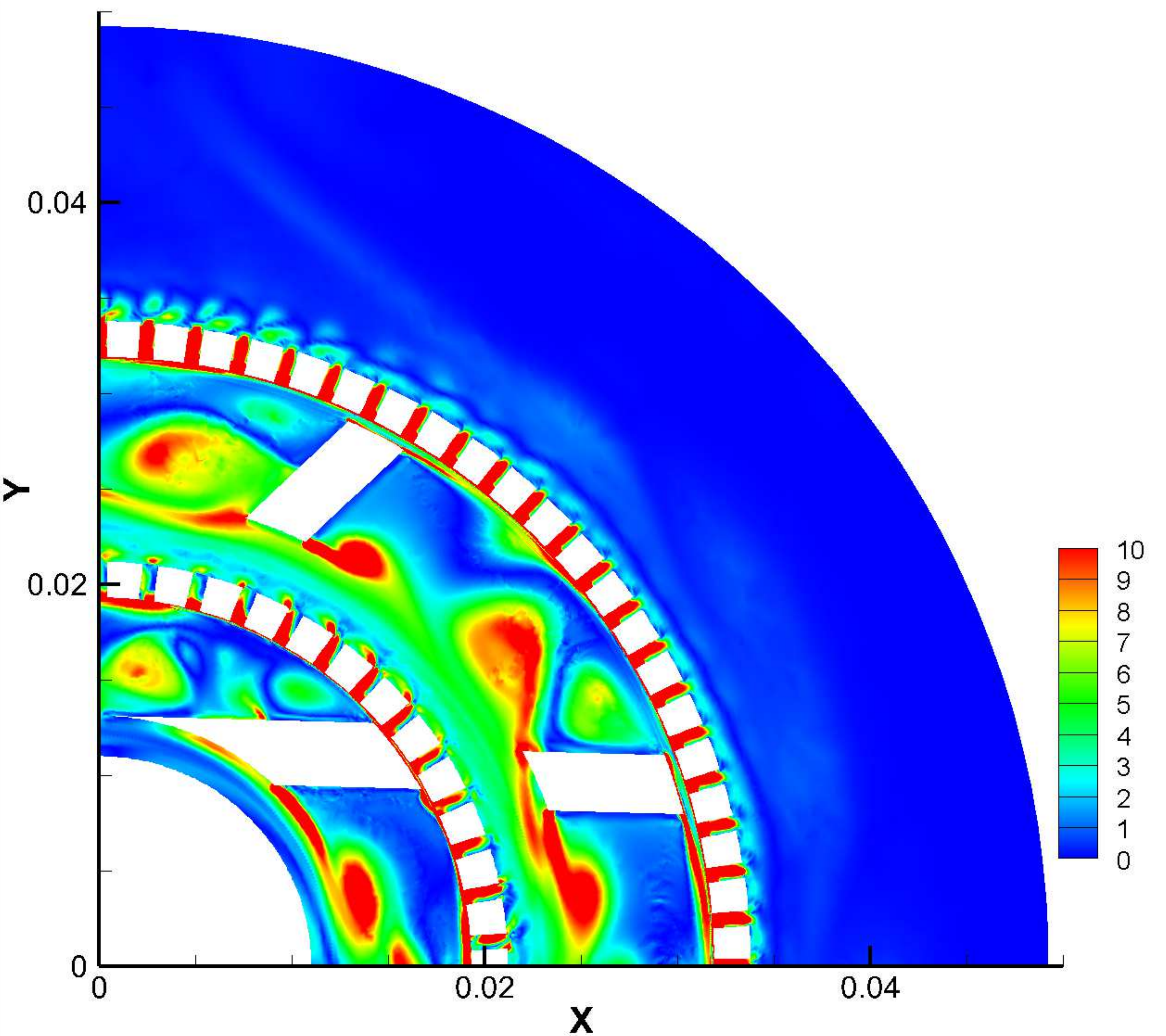}}

\centering\subfigure[$k-\omega$ SST at Q=3000 kg/hr]{\includegraphics[scale=0.13]{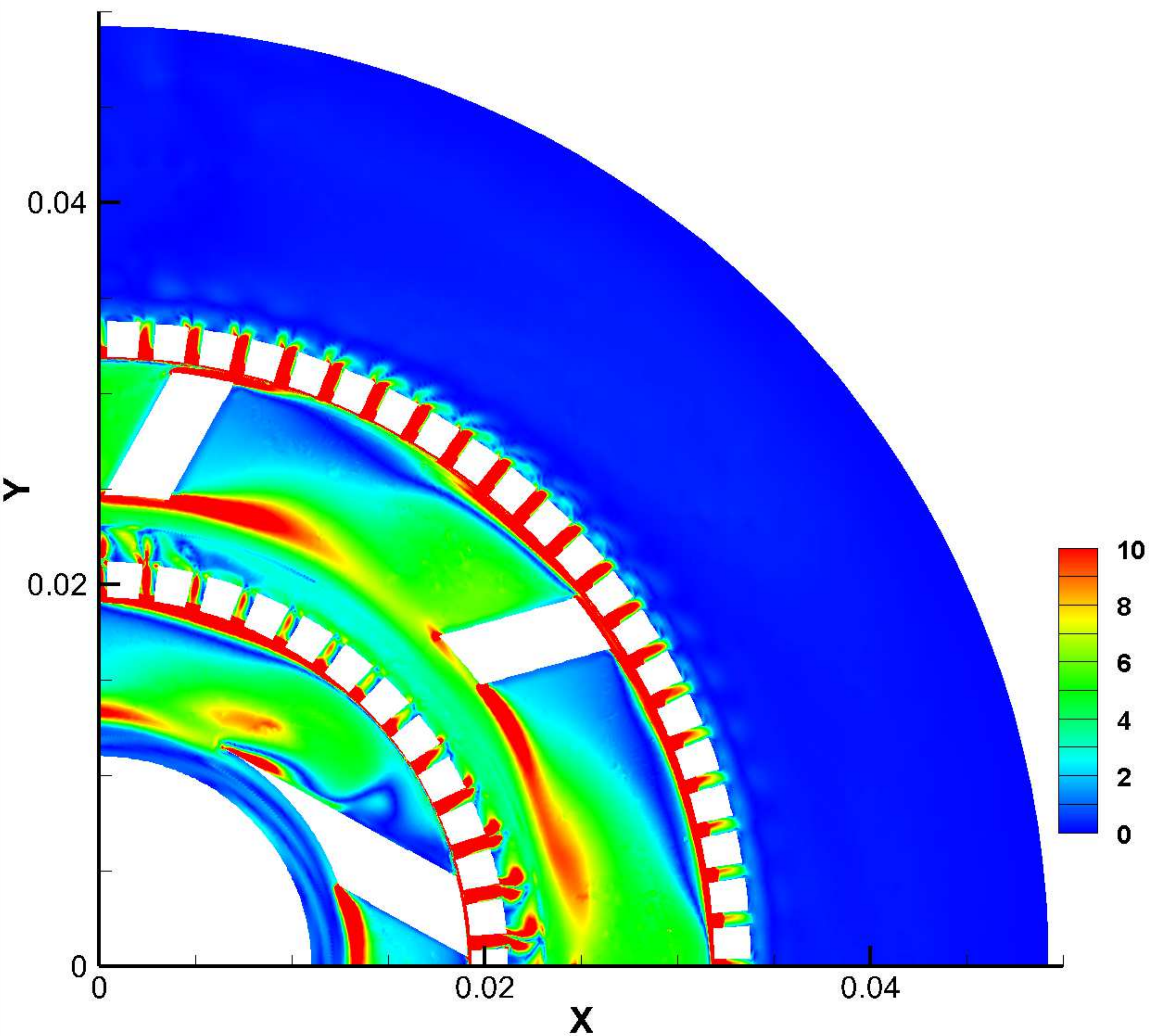}}\subfigure[SSG at Q=3000 kg/hr]{\includegraphics[scale=0.13]{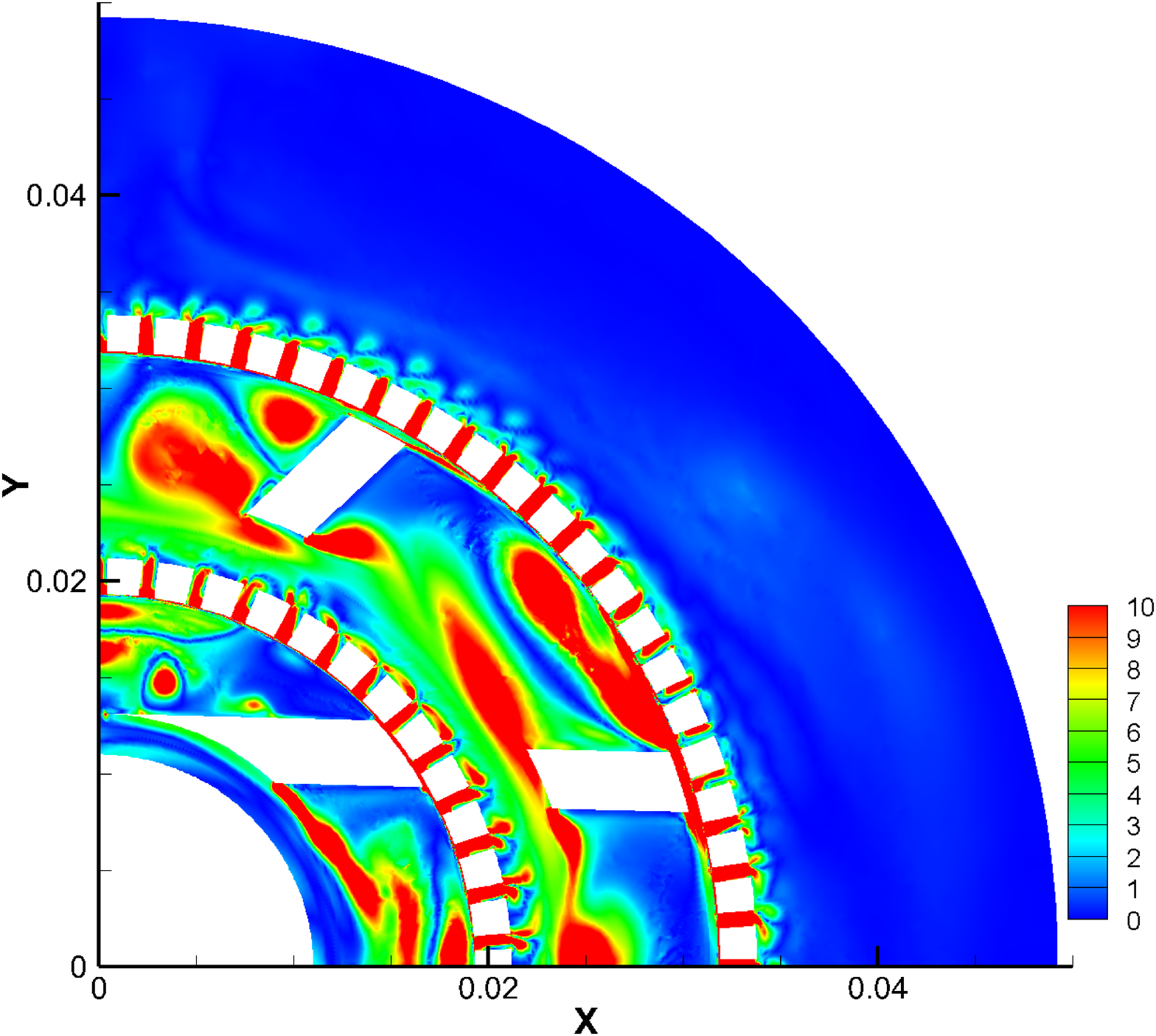}}

\caption{ Vorticity predictions from $k-\omega$ SST model and SSG model. The vorticity is normalised by the rotation speed of the mixer.\label{fig:Vort_SST_vs_SSG}}
\end{figure}

In order to investigate the differences in flow characterisation between the EVMs and RSM, the predicted velocity magnitude and vorticity magnitude are compared in \ref{fig:Vel_SST_vs_SSG} and \ref{fig:Vort_SST_vs_SSG} respectively. These results have been obtained using the sliding mesh method. The results from $k-\epsilon$ model are not shown here as they are similar to the results obtained from the $k-\omega$ SST model. There are subtle differences in the predicted velocity magnitude at different flow rates by the $k-\omega$ SST and SSG models as shown in \ref{fig:Vel_SST_vs_SSG}. \ref{fig:Vort_SST_vs_SSG} shows the vorticity field predicted by using different turbulence models at different flow rates. High regions of vorticity imply high velocity gradients promoting higher rate of mixing within the mixer. It can be seen in \ref{fig:Vort_SST_vs_SSG}b,d \& f that the SSG model predicts higher levels of vorticity at all flow rates when compared with the $k-\omega$ SST model predictions in \ref{fig:Vort_SST_vs_SSG}a,c \& e. 

\begin{figure}[htbp]
\centering\subfigure[SSG model]{\includegraphics[scale=0.1]{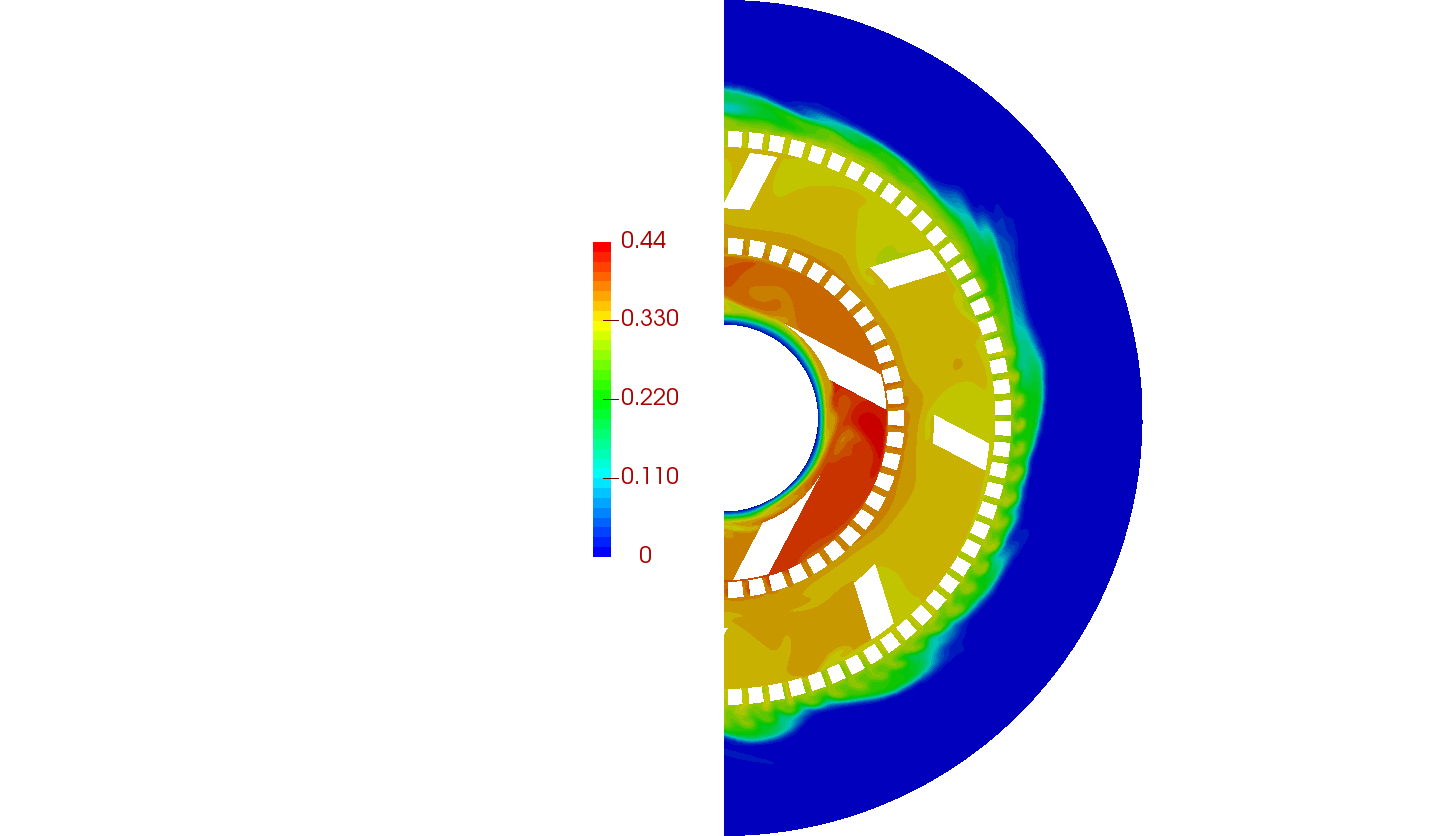}}\label{fig:Scalar_ssg}
\centering\subfigure[SST model]{\includegraphics[scale=0.1]{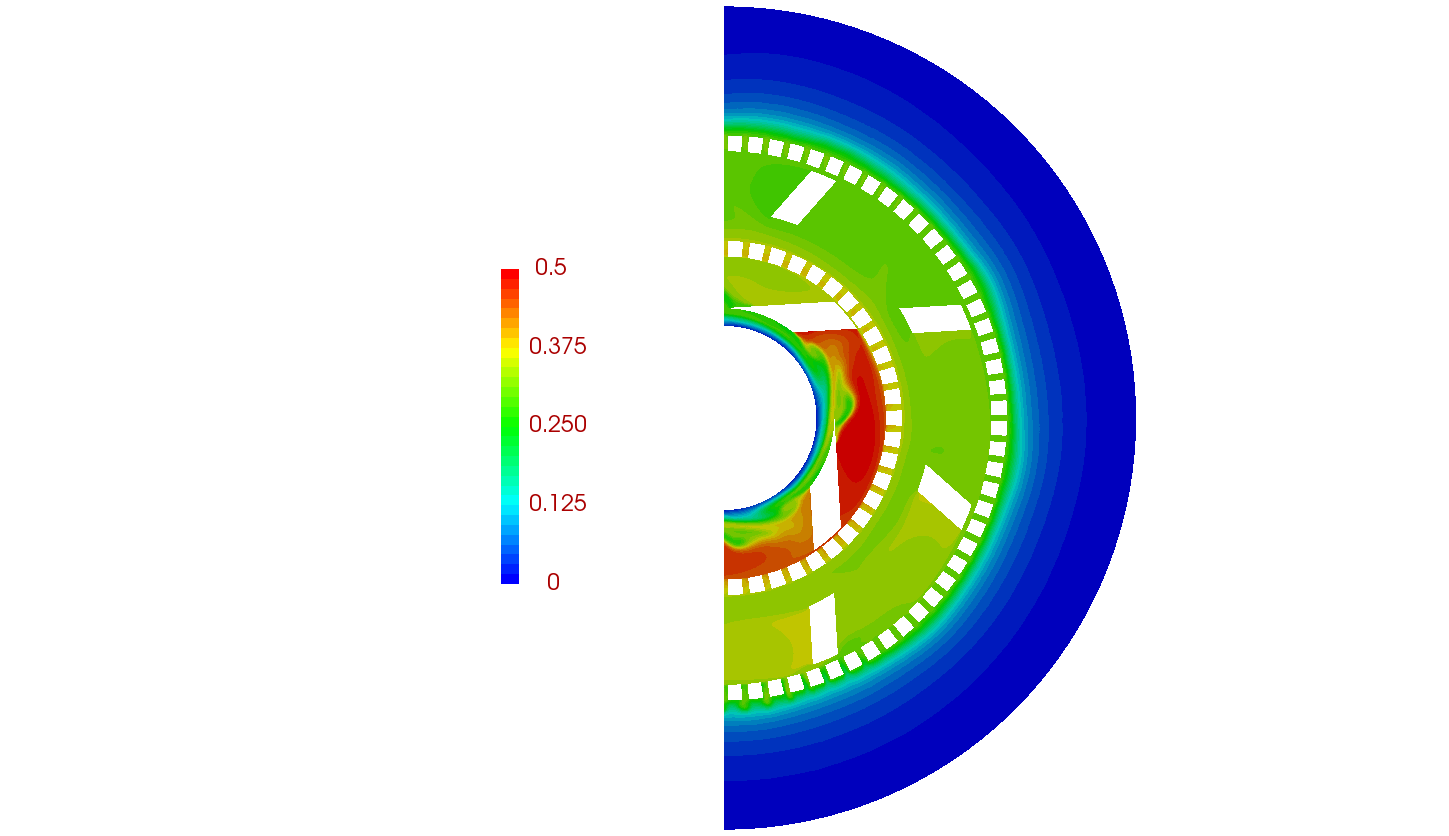}}\label{fig:Scalar_sst}
\caption{Distribution of scalar concentration after 10 revolutions using different turbulence models. Mixer is rotating at 6000 rpm with zero inflow. Sliding mesh method is used.} \label{fig:Scalar_t}
\end{figure}

The influence of turbulence models on the predictions of scalar mixing is investigated in more detail. The closure problem arising from Reynolds averaging the transport equations for the evolution of a passive scalar such as concentration $Y$ requires modelling of the turbulent flux $\overline{u_i'Y}$. In EVMs this is approximated as 
\begin{equation}
\overline{u_i'Y} \propto -\Gamma_t \frac{\partial Y}{\partial x_i},
\end{equation}
where the turbulent diffusivity $\Gamma_t$ is approximated using the eddy viscosity $\mu_t$ and is thus a scalar quantity (scalar gradient diffusion hypothesis, SGDH). With this model the turbulent flux is aligned with the mean scalar gradient. In RSMs, this model may be generalised to obtain 
\begin{equation}
\overline{u_i'Y} \propto -\frac{k}{\epsilon}\overline{u_i'u_j'}\frac{\partial Y}{\partial x_j},  
\end{equation}
which is known as the generalised gradient diffusion hypothesis (GGDH). The turbulent diffusivity in this model is a tensor which is an improvement over the SGDH as it allows anisotropy into the scalar flux model and coupling of the scalar flux with the Reynolds stresses. The distribution of the concentration field from its initial condition (\ref{fig:Scalar_m}a)  predicted using the SSG model (\ref{fig:Scalar_t}a) is compared to the predictions using SST model (\ref{fig:Scalar_t}b). The results demonstrate that mixing occurring through turbulent diffusion mechanism is better predicted by the SSG model. 

\begin{figure}[htbp]
\centering\subfigure[Normalised $k$ using SSG model]{\includegraphics[scale=0.1]{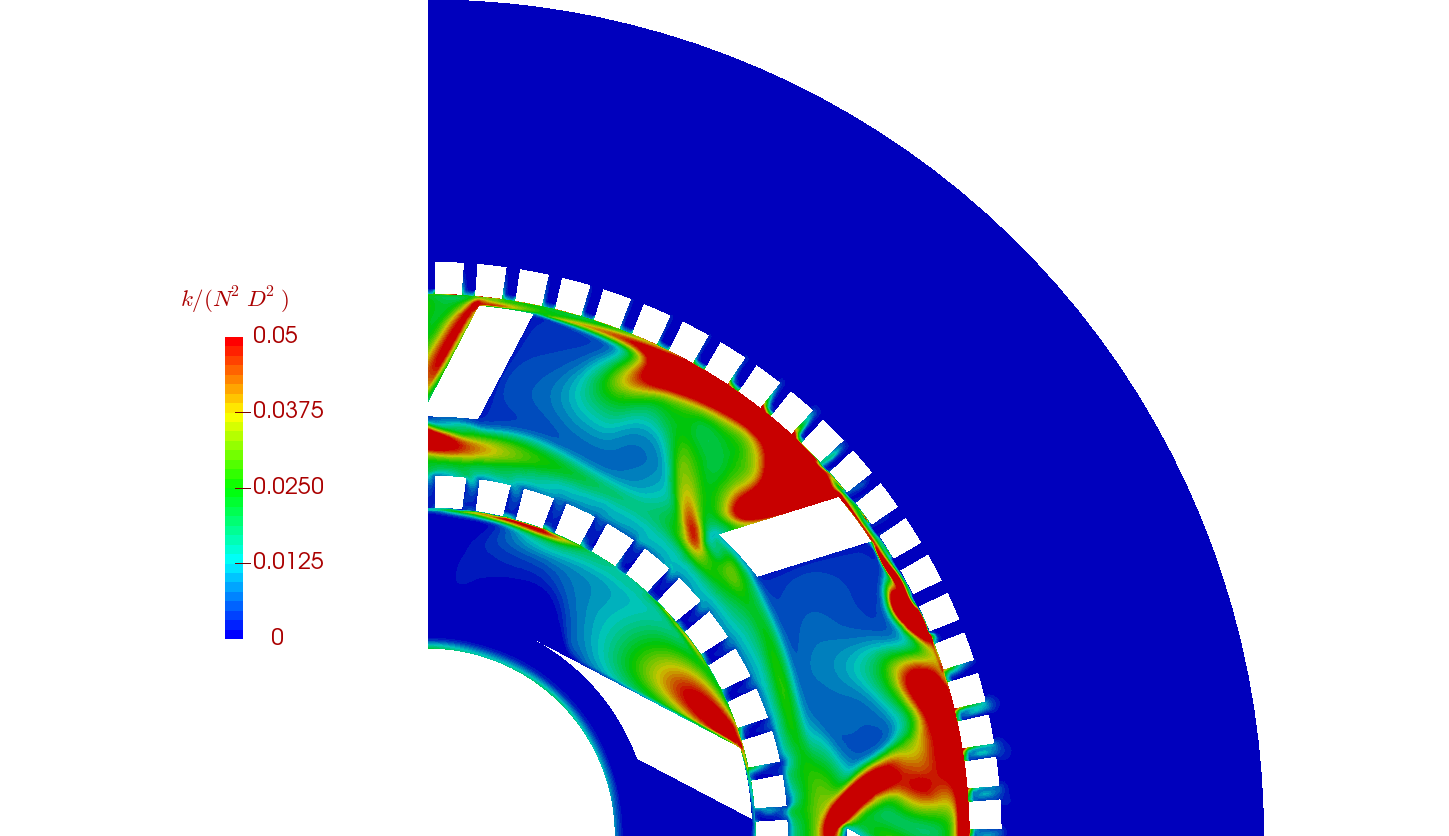}}
\centering\subfigure[Normalised $\epsilon$ using SSG model]{\includegraphics[scale=0.1]{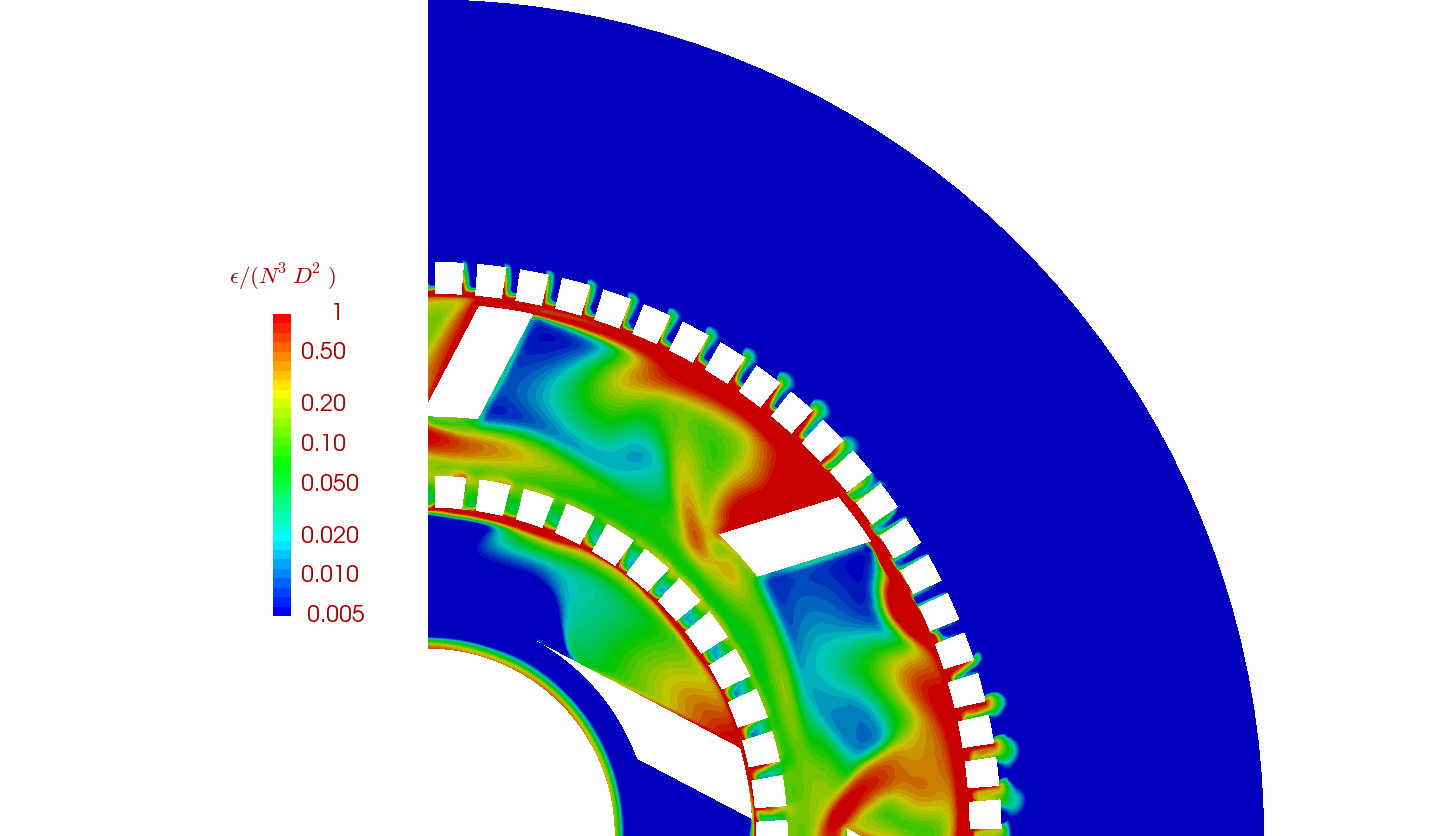}}\\
\centering\subfigure[Normalised $k$ using SST model]{\includegraphics[scale=0.1]{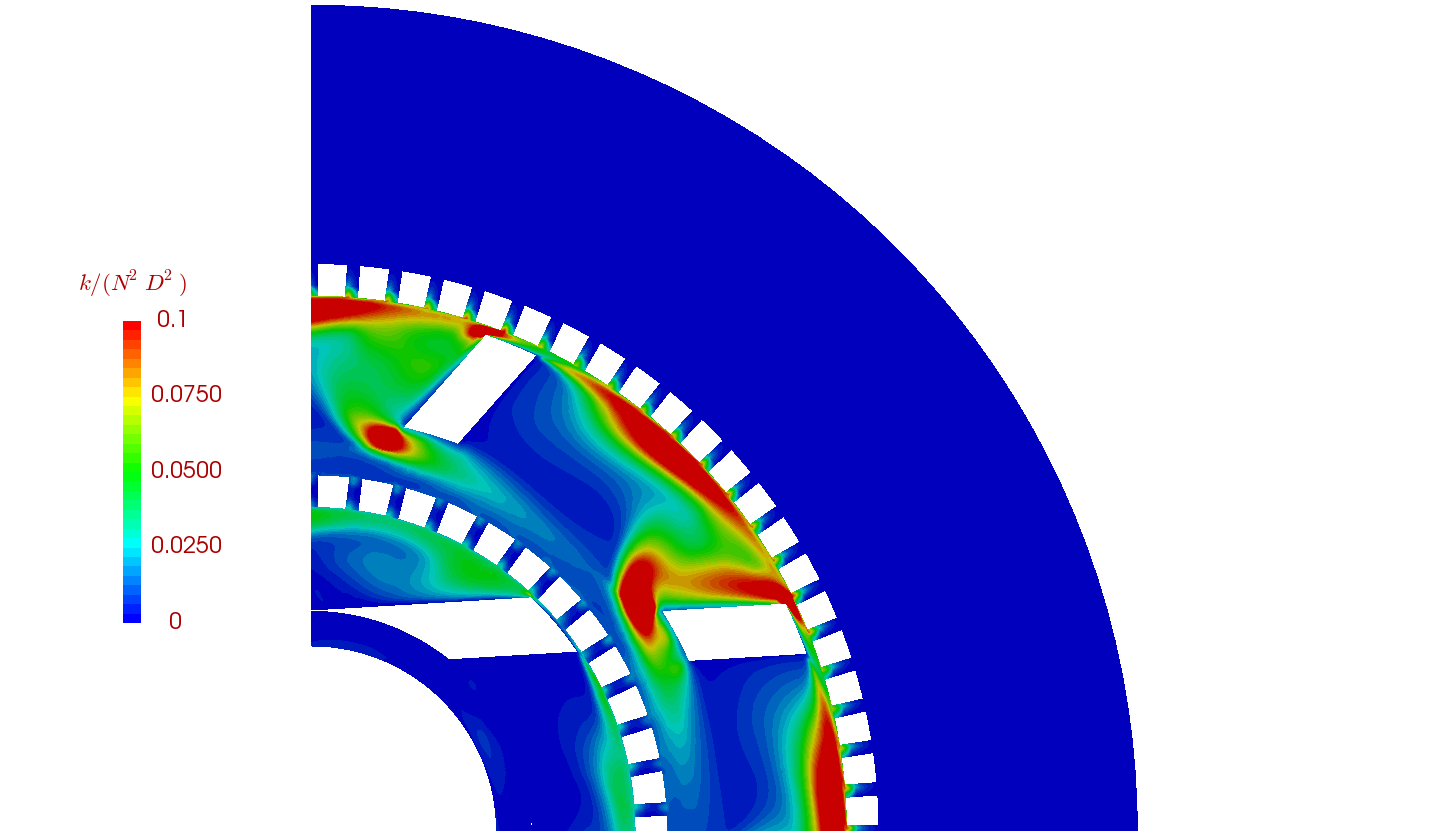}}
\centering\subfigure[Normalised $\epsilon$ using SST model]{\includegraphics[scale=0.1]{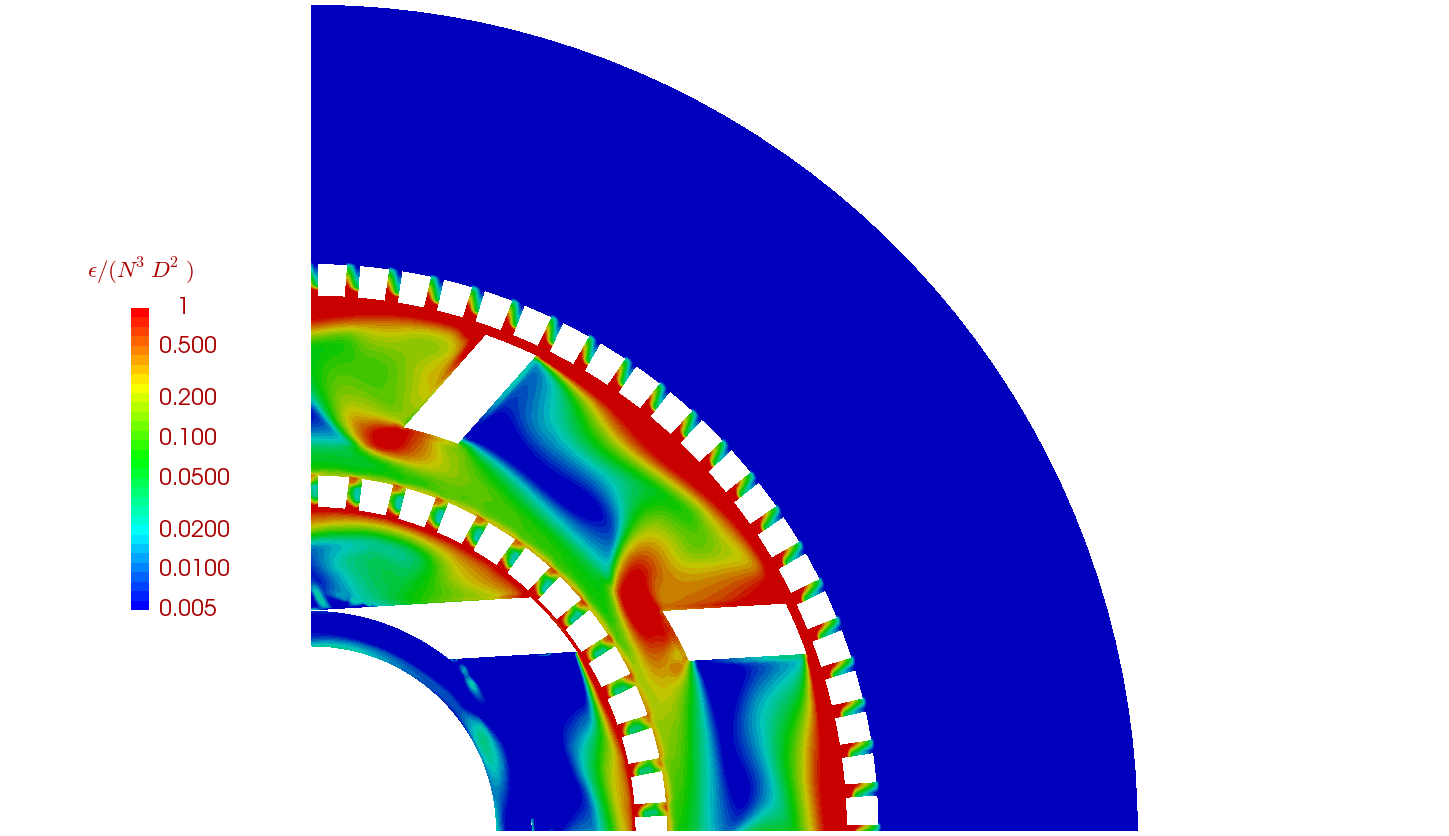}}\\
\centering\subfigure[Normalised $k$ using $k-\epsilon$ model]{\includegraphics[scale=0.1]{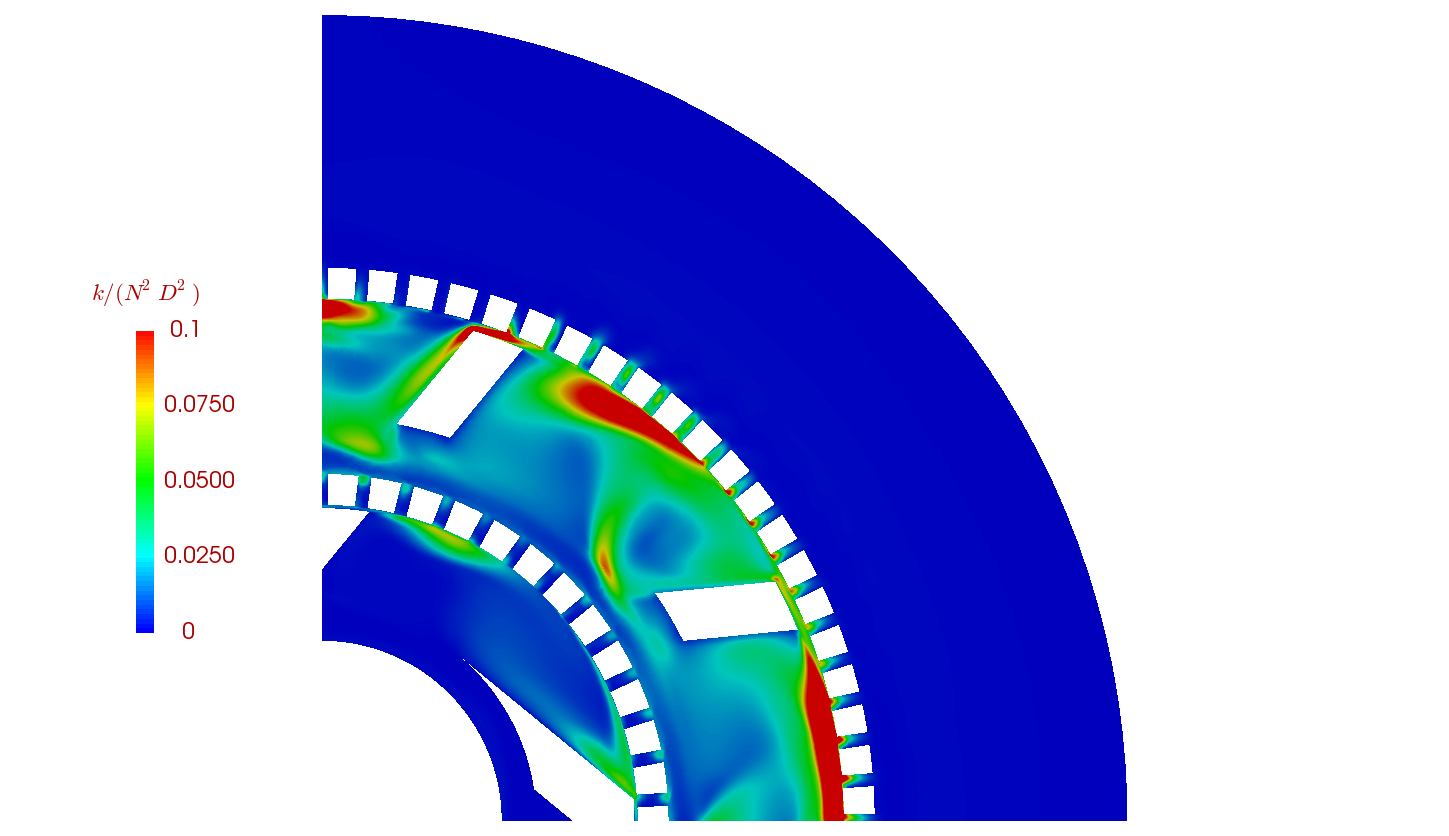}}
\centering\subfigure[Normalised $\epsilon$ using $k-\epsilon$ model]{\includegraphics[scale=0.1]{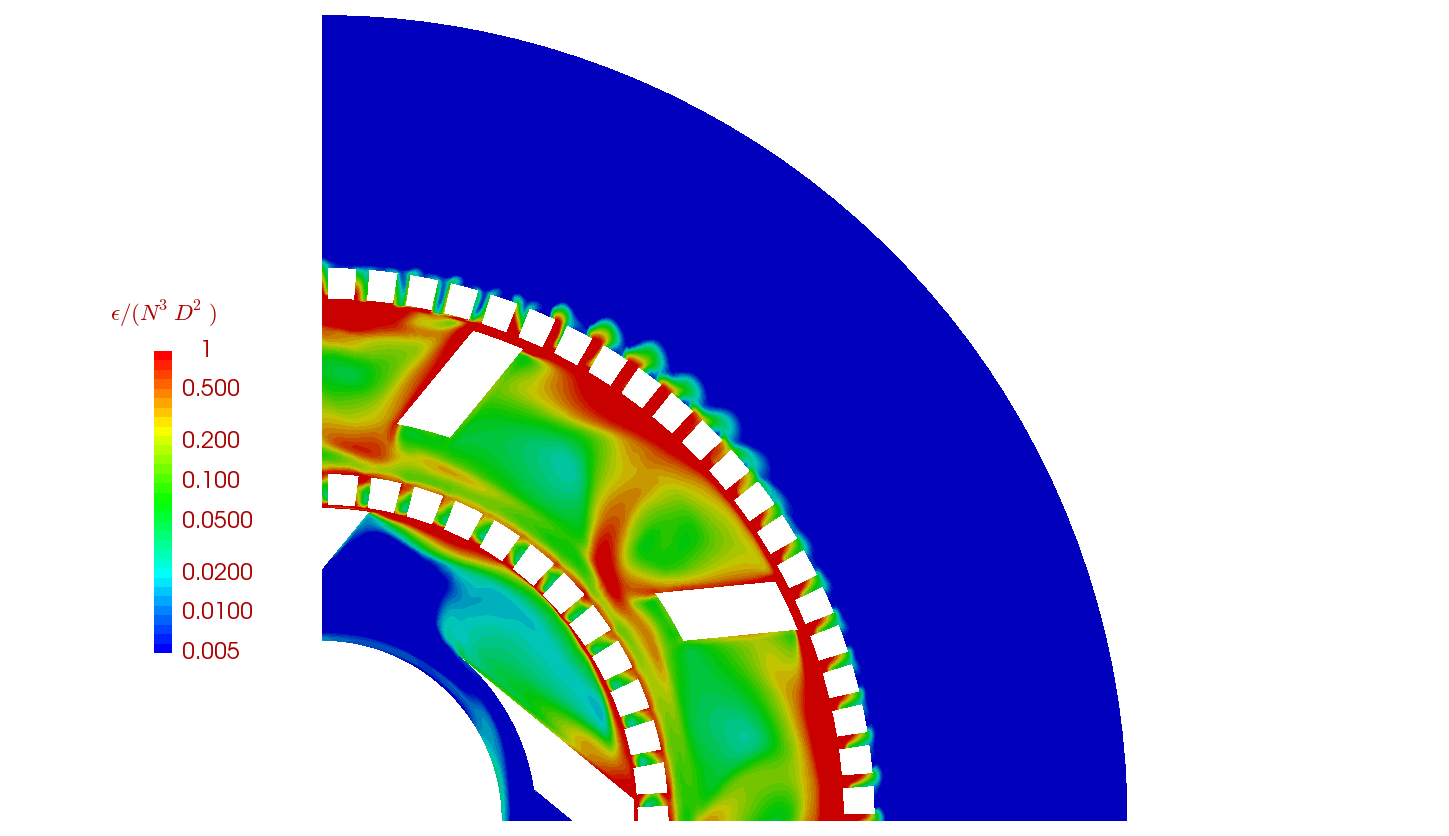}}
 \caption{Prediction of turbulence quantities using different turbulence models and sliding mesh method. Mixer is rotating at 6000 rpm with zero flow rate}\label{fig:turbke}
\end{figure}

The prediction of turbulent kinetic energy $k$ and turbulent energy dissipation $\epsilon$ by the different turbulence models are shown in \ref{fig:turbke}. The salient features of the turbulent kinetic energy field, peak regions on the pressure side of the blade and low regions on the suction side, are predicted by the SSG (\ref{fig:turbke}a) and SST models (\ref{fig:turbke}c). The $k-\epsilon$ model (\ref{fig:turbke}e) performs comparatively poorly in this regard. Turbulent energy dissipation is highest in the high shear regions around the stator screens and on the pressure side of the rotor blades. The $k-\epsilon$ model (\ref{fig:turbke}f) is seen to predict higher dissipation compared to the other models but predicts dissipation occurring in flows emanating from the outer screen which is also predicted by the SSG model (\ref{fig:turbke}b) but not the SST model (\ref{fig:turbke}d). 

\begin{figure}[htbp]
\centering
\includegraphics[scale=0.3]{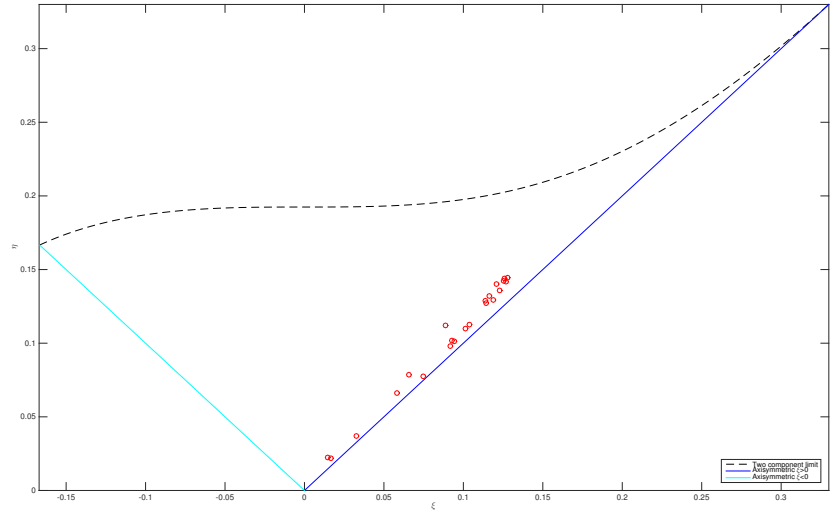}
\caption{Invariants $\xi$ and $\eta$ of the Reynolds stress anisotropy tensor $b_{ij}$. Special states of the tensor are indicated through the Lumley triangle \cite{Pope2000}. Symbols \textcolor{red}{o} indicate values of the invariants along x and y axis on the plane of the mixer}
\label{fig:lumley}
\end{figure}

\begin{figure}[htbp]
\centering
\subfigure[$\eta$]{\centering\includegraphics[scale=0.1]{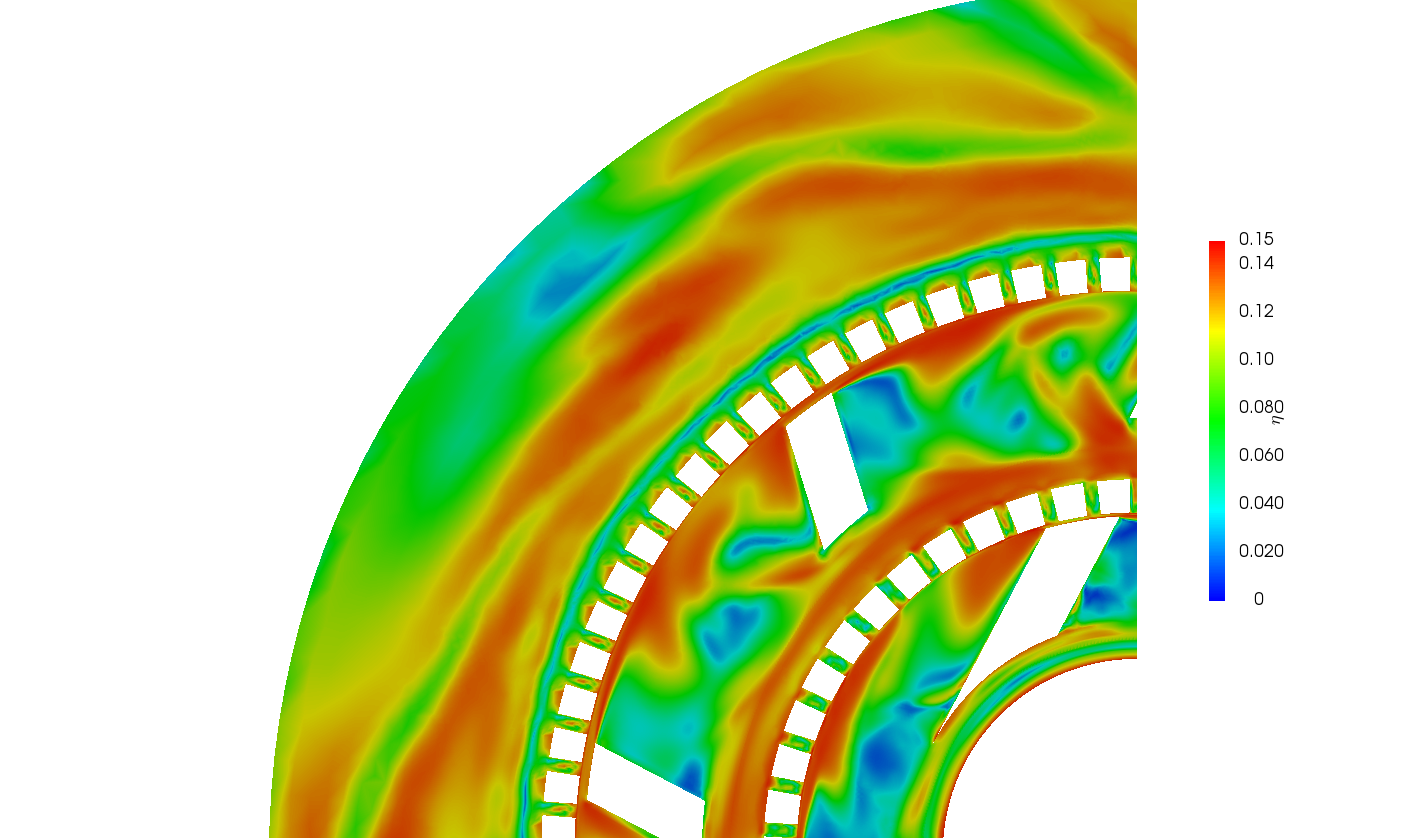}}
\subfigure[$\xi$]{\centering\includegraphics[scale=0.1]{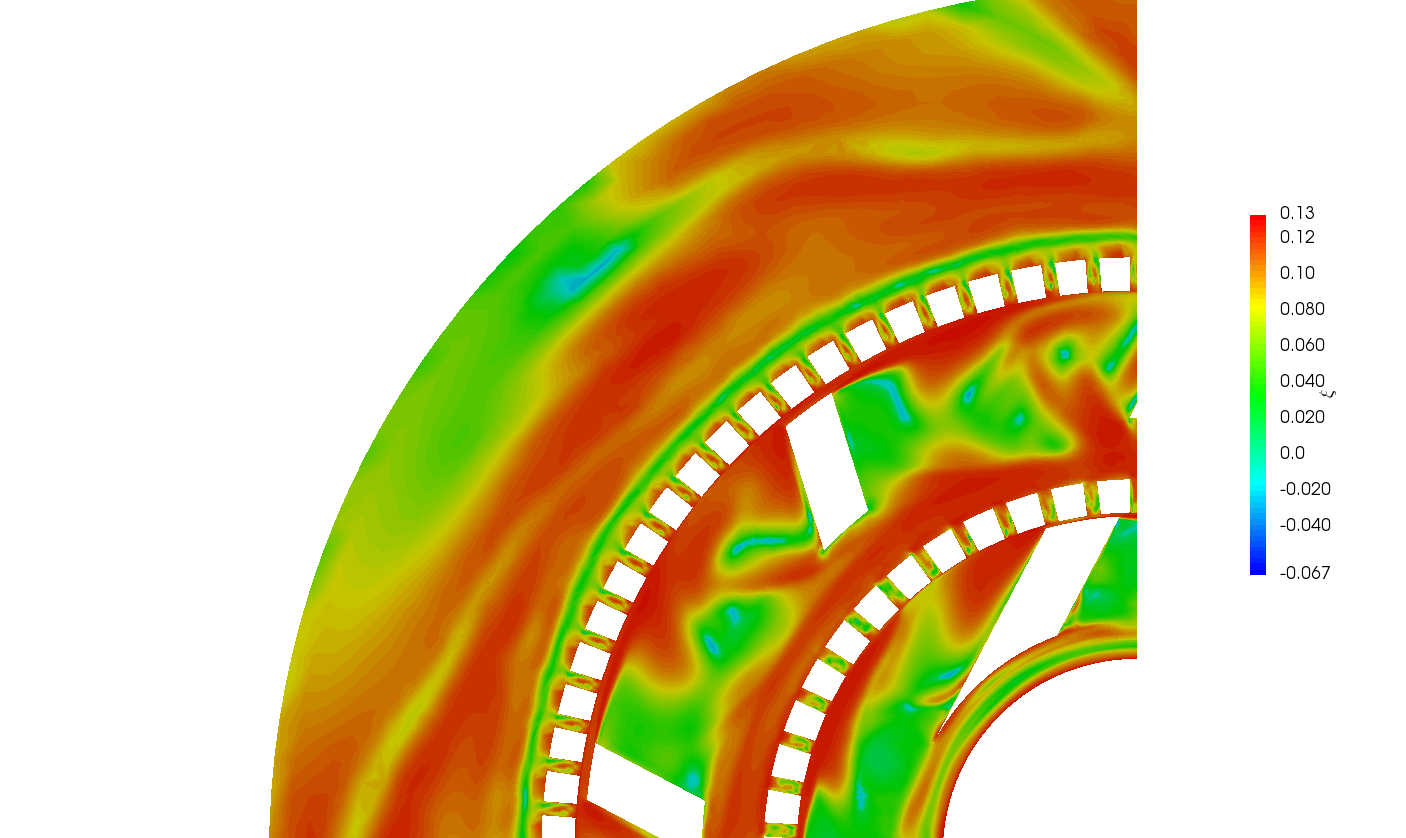}}
\caption{Second and third invariants of the Reynolds anisotropy tensor in mixer at 6000rpm and zero flow rate obtained using sliding mesh method. \label{fig:Second-and-third_invariants}}
\end{figure}

 The overall distribution of turbulent stresses in the mixer and hence the suitability of an EVM or RSM model can be obtained by examining the normalised Reynolds stress anisotropy tensor $b_{ij}$ defined as:
\begin{equation}
b_{ij}=\frac{\overline{u_{i}^{'}u_{j}^{'}}}{\overline{u_{k}^{'}u_{k}^{'}}}-\frac{1}{3}\delta_{ij}.\label{eq:anisotropy_tensor}
\end{equation}
Using \ref{eq:anisotropy_tensor} it can be seen that the anisotropy tensor has zero trace. Hence its first principal invariant
\begin{equation}
I_b = b_{ii} = 0.
\end{equation}
The state of anisotropy of the turbulent stresses can thus be investigated using its two remaining independent principal invariants. These invariants are defined as:
\begin{equation}
II_b = -\frac{1}{2}b^2_{ii},\label{eq:iib}
\end{equation}
\begin{equation} 
III_b = \frac{1}{3}b^3_{ii}.\label{eq:iib}
\end{equation}
On evaluating the principal invariants in principal axes, the second invariant defines the normal distance of the deviatoric stress plane from the isotropic vector and together with the third invariant fixes precisely the stress state on this plane. Pope \cite{Pope2000} proposes a simpler graphical representation of the anisotropic state of the Reynolds stresses in a turbulent flow using a $\xi-\eta$ plane, where
\begin{equation}
6\eta^{2}=-2II_{b}=b_{ii}^{2},\label{eq:second_invariant}
\end{equation}
and
\begin{equation}
6\xi^{3}=3III_{b}=b_{ii}^{3}.\label{eq:third_invariant}
\end{equation}
Analysing these invariants allows the turbulent state to be characterised via the Lumley triangle \cite{Pope2000} and to identify strongly anisotropic behaviour where EVMs would provide particularly poor predictions. Special states of anisotropy of the Reynolds stress tensor are indicated through lines on the Lumley triangle (\ref{fig:lumley}).
The turbulent stresses are fully isotropic wherever $\eta$ and $\xi$ are equal to zero with non-zero values implying anisotropic behaviour. \ref{fig:Second-and-third_invariants} shows the distribution of $\eta$ and $\xi$ determined from the Reynolds stresses calculated using the SSG model and the sliding mesh method. It can be seen that $\eta$ takes positive values throughout the mixer especially in narrow passages and close regions of the stator screens. $\xi$ is positive throughout the domain and is close to $\eta$ in magnitude. From the Lumley triangle (\ref{fig:lumley}) this indicates that the turbulent stresses are axisymmetric in these regions. Regions of isotropic turbulence can be observed on suction side of the blades of the rotors. This return to isotropy as the turbulence decays in these wake regions is captured by the SSG model. These features of the turbulence will not be predicted by the EVMs and a RSM model is needed to accurately predict the hydrodynamics of the mixer.  

\subsubsection{Power predictions}
\begin{figure}[htbp]
\centering
\includegraphics[scale=0.45]{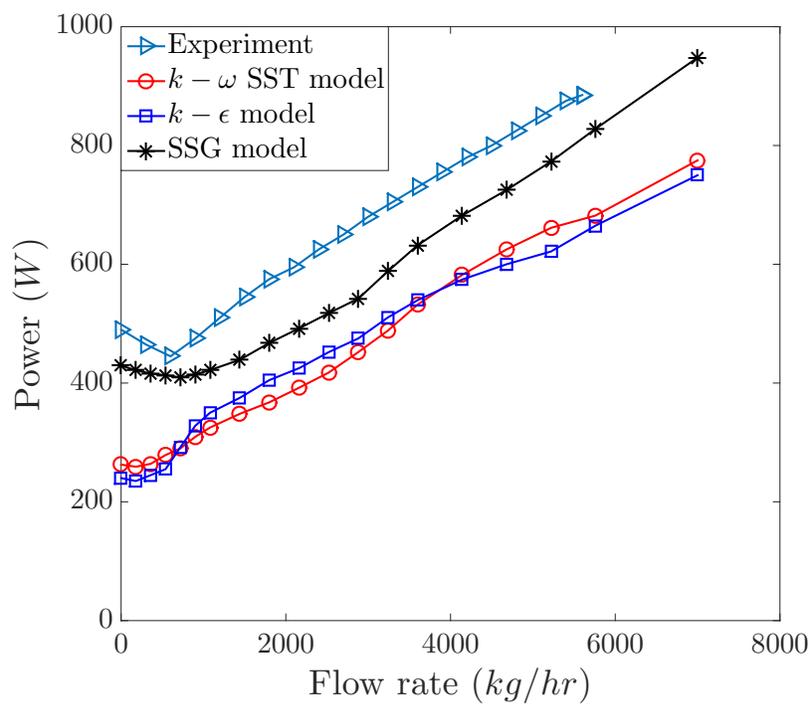}
 \caption{Power curve for Silverson 150/250 mixer using RSM and EVM turbulence models alongwith sliding mesh algorithm}
 \label{fig:EVM_vs_RSM}
\end{figure}

Power predictions from the eddy viscosity models ($k-\epsilon$ and $k-\omega$ SST) and the second moment closure model (SSG model) are presented in \ref{fig:EVM_vs_RSM}. The solutions have been obtained using the sliding mesh method. 
The biggest discrepancy in the power prediction from the EVMs is at low flow rates ($Q<1000$ kg/hr).
Note that across all flow rates the EVMs tend to predict a consistently lower power; the SSG model is much closer to the experimental data. The main reason for the better performance of the SSG model is due to the fact that the second moment closure models are able to capture the local anisotropy of the Reynolds stresses, thus leading to a better prediction of primary and secondary flows in the mixer. 

Cooke et al \cite{Cooke2011} proposes an expression to calculate power as :
 \begin{equation}
P=P_{O_{Z}}\rho N^{3}D^{5}+k_{1}QN^{2}D^{2},\label{eq:Power_prediction}
\end{equation}
where $P_{O_{Z}}$ is the power number at zero mass flow rate, $\rho$ is the density of the fluid, $D$ is the rotor diameter, $Q$ is the mass flow rate and $k_{1}$ is a proportionality constant. To evaluate the power using \ref{eq:Power_prediction}, the values for $P_{O_{Z}}$ and $k_{1}$ are required, and the simulations can be used to calculate these constants for a given Silverson mixer thereby reducing the requirement for physical plant trials. The calculated power vs flow rate data presented in \ref{fig:EVM_vs_RSM} is used to perform a linear fit, with the values for $P_{O_{Z}}$ and $k_{1}$ obtained from the y-axis intercept and slope respectively. The resulting values obtained using the different turbulence models are presented in table \ref{tab:Coefficients-for-power}. The RSM model (SSG) is able to predict $P_{O_{Z}}$ to within $12.5\%$ and $k_1$ to within $7.2\%$ of the experimental values. The $k-\omega$SST model is able to predict the slope $(k_1)$ with the same accuracy but underpredicts the power at zero flow-rate $(P_{O_{Z}})$. The $k-\epsilon$ model is the poorest performer among the three models in predicting the power constants. 
\begin{table}
\centering%
\begin{tabular}{|c|c|c|}
\hline 
 & $P_{O_{Z}}$ & $k_{1}$\tabularnewline
\hline 
Experiment & 0.475 & 7.611\tabularnewline
\hline 
$k-\epsilon$ model & 0.232 & 6.676\tabularnewline
\hline 
$k-\omega$ SST model & 0.254 & 7.069\tabularnewline
\hline 
SSG model & 0.416 & 7.061\tabularnewline
\hline 
\end{tabular}
\caption{Constants for power prediction obtained by using different turbulence models with the sliding mesh method.\label{tab:Coefficients-for-power} }
\end{table}

\subsection{Prediction of power number at different Reynolds numbers} 
\begin{figure}[htbp]
\centering\subfigure[Predicted power number at different Reynolds numbers by using the SSG model and sliding mesh method]{\includegraphics[scale=0.43]{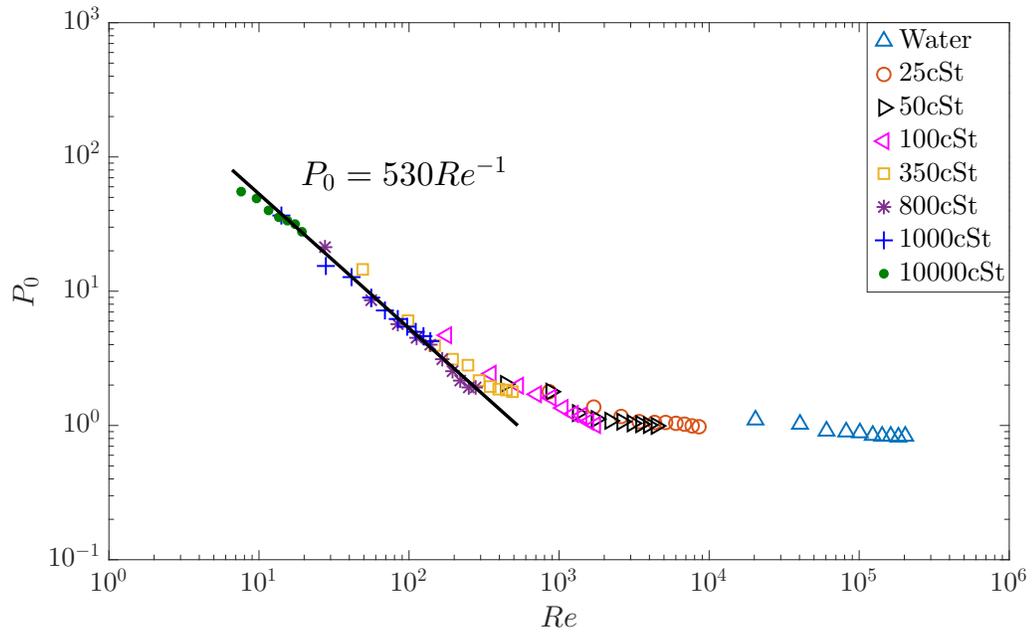}}
\centering\subfigure[Experimental power number at different Reynolds numbers \cite{Cooke2012}]{\includegraphics[scale=0.43]{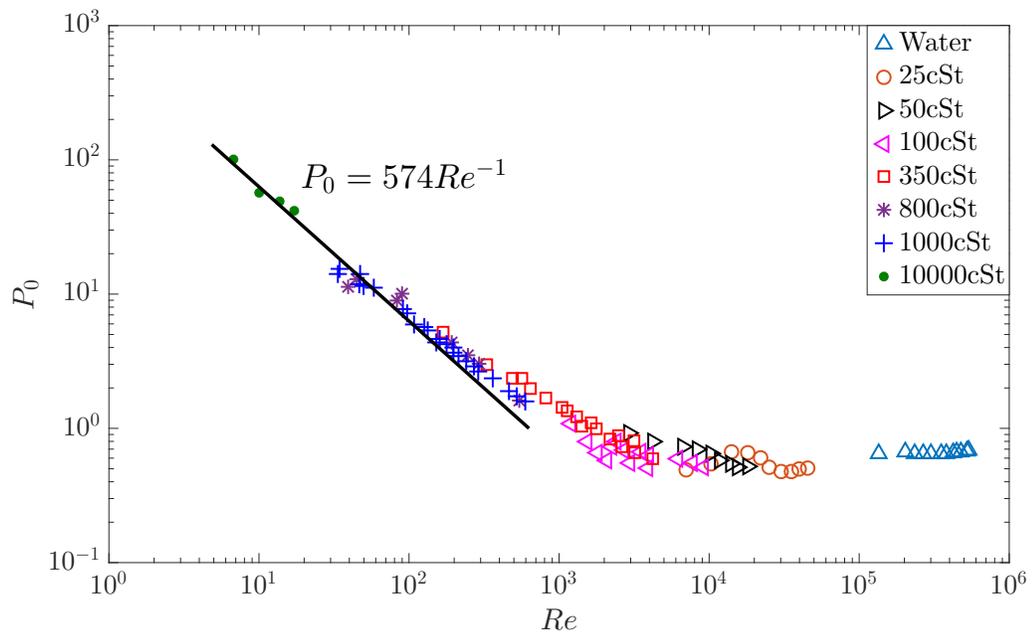}}
 \caption{Full power curve for Silverson 150/250 mixer.}
 \label{fig:Po_Re_curve}
\end{figure}

The power consumption of a mixer using a Newtonian fluid is usually expressed in the form of dimensionless power number $(P_{0})$ obtained by setting $k_1=0$ in \ref{eq:Power_prediction} \cite{Cooke2012}:
 \begin{equation}
P_{0}=\frac{P}{\rho N^{3}D^{5}}.\label{eq:Power_number}
\end{equation}
This expression provides a characteristic power curve that depends only on the swept diameter of the rotor and can be used to predict power requirements for any given fluid, rotor diameter, and rotational speed.
\ref{fig:Po_Re_curve}a shows the Reynolds number dependence of the power number as predicted using the SSG model and sliding mesh method. The Reynolds number in this case is defined as:  
 \begin{equation}
Re=\frac{\rho ND^{2}}{\mu},\label{eq:Reynolds_number}
\end{equation}
Note that the results at different Reynolds numbers for a given working fluid presented in \ref{fig:Po_Re_curve} are obtained by varying the rotor speed of the mixer. Linear variation of the power number with Reynolds number in the laminar regime and the invariance with Reynolds number in the turbulent regime are captured. The predicted power numbers are in good agreement with the experimental data of Cooke et al \cite{Cooke2012} as shown in \ref{fig:Po_Re_curve}b. 
The laminar power number is predicted to within $8\%$ and the turbulent power number to within $27\%$ of the experimental values. 
\section {Conclusions}
It has been shown in this paper that computational fluid dynamics (CFD) simulations are a valuable tool in understanding the hydrodynamics in high shear rotor-stator mixers, and can be used to calculate the constants required for the prediction of power in these mixers. A Silverson 150/250 MS in-line mixer is used as a representative configuration for this investigation. Comparisons between solution methods using a sliding mesh and multiple reference frame (MRF) algorithms are made. The sliding mesh method is better able to capture the hydrodynamics within the mixer, resulting in improved power predictions. The choice of turbulence model used in the simulations is found to be critical. Two different classes of turbulence models are compared; the eddy viscosity models ($k-\epsilon$ and $k-\omega$ SST models) and the second moment closure model (SSG model). It is shown that the SSG model is required to accurately capture the salient flow and mixing features within the mixer. This also results in improved prediction of the variation of power consumption with flow rate. CFD simulations have been conducted to capture the full characteristic power curve of the Silverson mixer and it is found that the second moment closure model coupled with the sliding mesh algorithm leads to results which are in good agreement with the experimental data. CFD simulations can therefore be a valuable tool for scale up calculations. 
 
\section*{Appendix}
\setcounter{secnumdepth}{0}
\subsection{Implementation of turbulence models}
\subsubsection{$k-\epsilon$ model}
The most common form of the model developed by Jones and Launder \citep{23}
is used here. The transport equations used in $k-\epsilon$ model
are:

\begin{equation}
\frac{\partial\overline{k}}{\partial t}+\overline{u_{i}}\frac{\partial\overline{k}}{\partial x_{i}}=\frac{\partial}{\partial x_{j}}\left[\left(\nu+\frac{\nu_{t}}{\sigma_{k}}\right)\frac{\partial\overline{k}}{\partial x_{j}}\right]+P_{k-\epsilon}-\overline{\epsilon},\label{eq:transport_eq_for_k}
\end{equation}

\begin{equation}
\frac{\partial\overline{\epsilon}}{\partial t}+\overline{u_{i}}\frac{\partial\overline{\epsilon}}{\partial x_{i}}=\frac{\partial}{\partial x_{i}}\left[\left(\nu+\frac{\nu_{t}}{\sigma_{\epsilon}}\right)\frac{\partial\overline{\epsilon}}{\partial x_{i}}\right]+C_{\epsilon1}\frac{\overline{\epsilon}}{\overline{k}}P_{k-\epsilon}-C_{\epsilon2}\frac{\overline{\epsilon}^{2}}{\overline{k}},\label{eq:transport_eq_for_epsilon_k-e_model}
\end{equation}
where 
\begin{equation}
P_{k-\epsilon}=-\overline{u_{i}^{'}u_{j}^{'}}\frac{\partial\widetilde{u}_{i}}{\partial x_{j}}.\label{eq:K-E_production}
\end{equation}
The turbulent viscosity $\mu_{t}$ is calculated as : 
\begin{equation}
\nu_{t}=C_{\mu}\frac{\overline{k}^{2}}{\overline{\epsilon}}\label{eq:eddy_viscosity_k-e model}
\end{equation}
The model constants $C_{\mu},C_{\epsilon1}\text{\,\ and\,\ }C_{\epsilon2}$
in eq.\eqref{eq:transport_eq_for_k} and eq.\eqref{eq:transport_eq_for_epsilon_k-e_model}are
given in table \ref{tab:The-values-of k-e model}.

\begin{table}[!htb]
\centering%
\begin{tabular}{|c|c|c|c|c|}
\hline 
$C_{\mu}$ & $\sigma_{k}$ & $\sigma_{\epsilon}$ & %
$C_{\epsilon1}$%
 & %
$C_{\epsilon2}$%
\tabularnewline
\hline 
0.09 & 1.0 & 1.3 & 1.44 & 1.92\tabularnewline
\hline 
\end{tabular}

\caption{\label{tab:The-values-of k-e model}Values of the empirical constants
in the $k-\epsilon$ model }
\end{table}
\subsubsection{$k-\omega$ SST model}
The standard $k-\omega$ SST model proposed by Menter \citep{Menter1994}
is also used for comparison. It blends the $k-\omega$ formulation
in the boundary layer and the free stream independence of the $k-\epsilon$
model in the far field. The governing equations for $k-\omega$ SST
model are :

\begin{equation}
\frac{\partial\overline{k}}{\partial t}+\overline{u}_{i}\frac{\partial\overline{k}}{\partial x_{j}}=\frac{\partial}{\partial x_{j}}\left[\left(\nu+\frac{\nu_{t}}{\sigma_{k}}\right)\frac{\partial\overline{k}}{\partial x_{j}}\right]+P_{k-\omega}-\beta^{*}\overline{\omega}\overline{k},\label{eq:SST_k_transport_equation}
\end{equation}

\begin{align}
\frac{\partial\overline{\omega}}{\partial t}+\frac{\partial\overline{u}_{j}\overline{\omega}}{\partial x_{j}} & =\frac{\partial}{\partial x_{j}}\left[\left(\nu+\frac{\nu_{t}}{\sigma_{\omega}}\right)\frac{\partial\overline{\omega}}{\partial x_{j}}\right]+\gamma\left\Vert \overline{S}\right\Vert ^{2}-\beta\overline{\omega}^{2}\nonumber \\
 & +2\left(1-F_{1}\right)\frac{1}{\sigma_{\omega_{2}}\overline{\omega}}\frac{\partial\overline{k}}{\partial x_{j}}\frac{\partial\overline{\omega}}{\partial x_{j}},\label{eq:SST_omega_transport_eq}
\end{align}
where 
\begin{equation}
P_{k-\omega}=min\left(-\overline{u_{i}^{'}u_{j}^{'}}\frac{\partial\overline{u}_{i}}{\partial x_{j}},10\beta^{*}\overline{k}\overline{\omega}\right).\label{eq:Production_kW}
\end{equation}
Any coefficient $\alpha$ in this model is calculated from 
\begin{equation}
\alpha=F_{1}\alpha_{1}+\left(1-F_{1}\right)\alpha_{2},\label{eq:blending_of_coefficients}
\end{equation}
 where subscript $1$ corresponds to the coefficients in the $k-\omega$
model and subscript $2$ corresponds to the coefficients in the $k-\epsilon$
model. $F_{1}$ is the blending function in eq.\eqref{eq:SST_omega_transport_eq}
defined as:
\begin{equation}
F_{1}=tanh\left(arg_{1}^{4}\right),\label{eq:SST_blending_function}
\end{equation}
where 
\begin{equation}
arg_{1}=min\left[max\left(\frac{\sqrt{\overline{k}}}{\beta^{*}\overline{\omega}y};\frac{500\nu}{y^{2}\overline{\omega}}\right);\frac{4\overline{k}}{\sigma_{\omega_{2}}CD_{k\omega}y^{2}}\right]\label{eq:arg_in_the_blending_function}
\end{equation}

\begin{equation}
CD_{k\omega}=max\left(2\frac{1}{\sigma_{\omega_{2}}\overline{\omega}}\frac{\partial \overline{k}}{\partial x_{j}}\frac{\partial\overline{\omega}}{\partial x_{j}},10^{-20}\right).\label{eq:CD_inarg1_k_omega_SST}
\end{equation}
$y$ in eq.\eqref{eq:CD_inarg1_k_omega_SST} represents the the distance
to the nearest wall, and $CD_{k\omega}$ is the positive part of the
cross diffusion term \citep{Menter1994}. The eddy viscosity is calculated
as \citep{Menter1994}: 
\begin{equation}
\nu_{t}=\frac{\overline{k}a_{1}}{max\left(a_{1}\overline{\omega};\left\Vert S\right\Vert F_{2}\right)}\label{eq:eddy_viscosity_SST_model}
\end{equation}
where

\begin{equation}
\left\Vert \overline{S}\right\Vert =\sqrt{2S_{ij}S_{ij}}\label{eq:strain_invarient}
\end{equation}

\begin{equation}
F_{2}=tanh\left(arg_{2}^{2}\right)\label{eq:SST_blending_function_2}
\end{equation}

\begin{equation}
arg_{2}^{2}=max\left(\frac{2\sqrt{\overline{k}}}{\beta^{*}\overline{\omega}y};\frac{500\nu}{y^{2}\overline{\omega}}\right)\label{eq:arg2_in_the_blending_function}
\end{equation}
The model constants for the $k-\omega$ SST model are given in table\eqref{tab:Model-constants-for_k_omega_SST}.

\begin{table}[!htb]
\centering%
\begin{tabular}{|c|c|c|c|c|c|c|c|c|c|c|}
\hline 
$\sigma_{k_{1}}$ & $\sigma_{\omega_{1}}$ & $\beta_{1}$ & $a_{1}$ & $\beta^{*}$ & $\kappa$ & $\gamma_{1}$ & $\sigma_{k_{2}}$ & $\sigma_{\omega_{2}}$ & $\beta_{2}$ & $\gamma_{2}$\tabularnewline
\hline 
1.176 & 2.0 & 0.075 & 0.31 & 0.09 & 0.41 & $\frac{\beta_{1}}{\beta^{*}}-\frac{\kappa^{2}}{\sigma_{\omega1}\sqrt{\beta^{*}}}$ & 1.0 & 1.168 & 0.0828 & $\frac{\beta_{2}}{\beta^{*}}-\frac{\kappa^{2}}{\sigma_{\omega2}\sqrt{\beta^{*}}}$\tabularnewline
\hline 
\end{tabular}

\caption{Model constants for the $k-\omega$ SST model\label{tab:Model-constants-for_k_omega_SST}}
\end{table}

\subsubsection{SSG model}
The standard SSG model proposed by  Speziale et al \cite{Speziale1991} is used as the second moment closure model. This model uses six Reynolds stress transport equations and a turbulent dissipation transport equation. The governing equations for the model are :

\begin{equation}
\frac{\partial\overline{u_{i}^{'}u_{j}^{'}}}{\partial t}+\overline{u_{k}}\frac{\partial\overline{u_{i}^{'}u_{j}^{'}}}{\partial x_{k}}=D_{ij}+P_{ij}+\phi_{ij}-\epsilon_{ij},\label{eq:RSM_transport_Eq}
\end{equation}
where 
\begin{equation}
D_{ij}=\frac{\partial}{\partial x_{k}}\left[\nu\frac{\partial\overline{u_{i}^{'}u_{j}^{'}}}{\partial x_{k}}-C_{s}\frac{\overline{k}}{\overline{\epsilon}}\overline{u_{k}^{'}u_{l}^{'}}\frac{\partial\overline{u_{i}^{'}u_{j}^{'}}}{\partial x_{l}}\right]
\end{equation}
\begin{equation}
P_{ij}=-\overline{u_{i}^{'}u_{k}^{'}}\frac{\partial\overline{u_{j}}}{\partial x_{k}}-\overline{u_{j}^{'}u_{k}^{'}}\frac{\partial\overline{u_{i}}}{\partial x_{k}}
\end{equation}
\begin{eqnarray}
\phi_{ij} & = & -C_{1}\overline{\epsilon}\overline{b_{ij}}+C_{1}^{'}\overline{\epsilon}\left(\overline{b_{ik}}\ \overline{b_{kj}}-\frac{1}{3}\overline{b_{mn}}\ \overline{b_{nm}}\right)+C_{2}\overline{k}\ \overline{S_{ij}}\nonumber \\
 &  & +C_{3}\overline{k}\left(\overline{b_{ik}}\ \overline{S_{jk}}+\overline{b_{jk}}\ \overline{S_{ik}}-\frac{2}{3}\overline{b_{mn}}\ \overline{S_{mn}}\delta_{ij}\right)\nonumber \\
 &  & +C_{4}\overline{k}\left(\overline{b_{ik}}\ \overline{\Omega_{jk}}+\overline{b_{jk}}\ \overline{\Omega_{ik}}\right).\label{eq:Phi_ij_model}
\end{eqnarray}
$\overline{b_{ij}}$, $\overline{\Omega_{ij}}$ and $\overline{S_{ij}}$
in Eq. \ref{eq:Phi_ij_model} are defined as : 
\begin{equation}
\overline{b_{ij}}=\frac{\overline{a_{ij}}}{2\overline{k}}-\frac{1}{3}\delta_{ij},
\end{equation}
where $a_{ij}$ is the anisotropy tensor,
\begin{equation}
\overline{S_{ij}}=\frac{1}{2}\left(\frac{\partial\overline{u_{i}}}{\partial x_{j}}+\frac{\partial\overline{u_{j}}}{\partial x_{i}}\right),
\end{equation}
\begin{equation}
\overline{\Omega_{ij}}=\frac{1}{2}\left(\frac{\partial\overline{u_{i}}}{\partial x_{j}}-\frac{\partial\overline{u_{j}}}{\partial x_{i}}\right).
\end{equation}
$\epsilon_{ij}$ in Eq. \ref{eq:RSM_transport_Eq} is closed under
the isotropic assumption and the transport equation proposed by Hanjali\'c
and Launder \cite{Hanjalic1972} is used: 
\begin{equation}
\frac{\partial\overline{\epsilon}}{\partial t}+\overline{u_{k}}\frac{\partial\overline{\epsilon}}{\partial x_{k}}=\frac{\partial}{\partial x_{j}}\left(C_{\epsilon}\frac{\overline{k}}{\overline{\epsilon}}\overline{u_{i}^{'}u_{j}^{'}}\frac{\partial\overline{\epsilon}}{\partial x_{j}}\right)+C_{\epsilon1}\frac{P_{ii}\overline{\epsilon}}{2\overline{k}}-C_{\epsilon2}\frac{\overline{\epsilon}^{2}}{\overline{k}}.\label{eq:RSM_dissipation}
\end{equation}
The constants in the above equations are given in table \ref{tab:Coefficients-of-the_SSG_model}.

\begin{table}[!htb]
\centering%
\begin{tabular}{|c|c|c|c|c|c|c|}
\hline 
$C_{1}$ & $C_{1}^{'}$ & $C_{2}$ & $C_{3}$ & $C_{4}$ & $C_{\epsilon1}$ & $C_{\epsilon2}$\tabularnewline
\hline 
$3.4+1.8P_{ii}/2\epsilon$ & 4.2 & $0.8-1.3\left(b_{ij}b_{ij}\right)^{0.5}$ & 1.25 & 0.4 & 1.44 & 1.83\tabularnewline
\hline 
\end{tabular}

\caption{Coefficients of the SSG model \label{tab:Coefficients-of-the_SSG_model}}

\end{table}

\pagebreak
\section*{References}
\bibliography{silverson,library}

\end{document}